\documentclass{aastex631}
\usepackage{soul}

\shorttitle{Drift effects over the Sun's Polarity Reversal}

\shortauthors{Aslam et al.}
\graphicspath{{./}{figures-1/}}

\begin{document}

\title{Unfolding Drift Effects for Cosmic Rays over the Period of the Sun's Magnetic Field Reversal}

\correspondingauthor{Xi Luo}
\email{xi.luo@iat.cn}

\author[0000-0001-9521-3874]{O.P.M. Aslam}
\affiliation{Shandong Institute of Advanced Technology (SDIAT)\\
250100 Jinan, Shandong Province \\
People's Republic of China}

\author[0000-0002-4508-6042]{Xi Luo}
\affiliation{Shandong Institute of Advanced Technology (SDIAT)\\
250100 Jinan, Shandong Province \\
People's Republic of China}

\author[0000-0003-0793-7333]{M.S. Potgieter}
\affiliation{Shandong Institute of Advanced Technology (SDIAT)\\
250100 Jinan, Shandong Province \\
People's Republic of China}
\affiliation{Institute for Experimental and Applied Physics (IEAP)\\
Christian-Albrechts-University in Kiel\\
24118 Kiel, Germany}

\author[0000-0001-5844-3419]{M.D. Ngobeni}
\affiliation{Centre for Space Research, North-West University \\
2520 Potchefstroom, South Africa}
\affiliation{School of Physical and Chemical Sciences, North-West University\\
2735 Mmabatho, South Africa}

\author[0000-0002-7723-5743]{Xiaojian Song}
\affiliation{Shandong Institute of Advanced Technology (SDIAT)\\
250100 Jinan, Shandong Province \\
People's Republic of China}




\begin{abstract}
A well-established, comprehensive 3-D numerical modulation model is applied to simulate galactic protons, electrons and positrons from May 2011 to May 2015, including the solar magnetic polarity reversal of Solar Cycle 24.
The objective is to evaluate how these simulations compare with corresponding AMS observations for 1.0-3.0 GV, and what underlying physics follows from this comparison in order to improve our understanding on how the major physical modulation processes change, especially particle drift, from a negative to a positive magnetic polarity cycle.
Apart from their local interstellar spectra, electrons and positrons differ only in their drift patterns, but they differ with protons in other ways such as their adiabatic energy changes at lower rigidity.
In order to complete the simulations for oppositely charged particles, antiproton modeling results are obtained as well.
Together, the observations and the corresponding modeling indicate the difference in the drift pattern before and after the recent polarity reversal and clarify to a large extent the phenomenon of charge-sign dependence during this period.   
The effect of global particle drift became negligible during this period of no well-defined magnetic polarity. 
The resulting low values of all particles' MFPs during the polarity reversal contrast their large values during solar minimum activity, and as such expose the relative contributions and effects of the different modulation processes from solar minimum to maximum activity. We find that the drift scale starts recovering just after the polarity reversal, but the MFPs keep decreasing or remain unchanged for some period after the polarity reversal. 

\end{abstract}



\section{Introduction} \label{sec:intro}

Galactic cosmic rays  (GCRs) are modulated by the heliospheric magnetic field (HMF) embedded in the solar wind. The level of modulation of these charged particles varies with time according to the level of solar activity and depends on the type of GCRs, their rigidity and where inside the heliosphere they are observed.
The transport of these GCRs is a superposition of different coherent physical processes: a) Convection caused by the outward directed solar wind velocity. b) Spatial diffusion caused by the scattering of these particles from magnetic field irregularities. c) Adiabatic energy losses. d) Particle drift caused by gradients, curvatures and the wavy heliospheric current sheet (HCS). In this context, see reviews by \citet{1984SSRv...37..201Q,2013SSRv..176..391K,2013SSRv..176..299M,2013SSRv..176..165P,2013LRSP...10....3P} and a recent comprehensive overview of the theory as applicable to the heliosphere by \citet{2022SSRv..218...33E}. 

The GCR intensity varies with different time scales, periodically usually on the longer time scales (months, 11 years and 22 years) and non-periodically usually on a shorter time scale (Forbush Decreases, etc.). This is determined by the solar activity cycle, the level of activity and the corresponding heliospheric conditions from the Sun to the heliospheric outer boundary. See also the recent observational review by \citet[][and references there-in]{2022SSRv..218...42R}.

All four major physical processes are considered to be important in modulating GCRs but their relative importance vary depending on the level of solar activity and the consequent modulation conditions. It is expected that particle drift plays a relatively dominant role during low and moderate solar activity conditions, depending on the direction of the HMF, while diffusion becomes dominant during high solar activity periods. In this context, see reviews by \citet{2013SSRv..176..265H, 2014AdSpR..53.1415P, 2017AdSpR..60..848P}, and also recent modeling by \citet{2019ApJ...873...70A,2021ApJ...909..215A}. Adiabatic energy changes are important throughout the solar cycle, and in fact is dominating the spectral shape of galactic protons, antiprotons and other light and heavier nuclei at lower rigidity, but not in case of galactic electrons and positrons at low rigidity; see illustrations by \citet{1982Ap&SS..84..519M} and \citet{2011JGRA..11612105S}. 

The focus here is on the long-term modulation cycles, with emphasis on how the intensity of GCRs varies from solar minimum to maximum activity conditions and especially what happens during the time of the reversal of the direction of the HMF, known as the polarity reversal, that always occurs around maximum solar activity. This is done by comparing our numerical simulations to available (published) GCR observations. 
Concerning the unfolding of drift effects and understanding how the major physical modulation processes for GCRs are changing from a negative to positive polarity through a period of uncertain polarity for Solar Cycle 24, we utilize proton, electron and positron observations averaged over Bartels rotations (BRs) from AMS \citep{2018PhRvL.121e1101A, 2018PhRvL.121e1102A} from May 2011 to May 2015 (BRs number 2426 - 2480).
The magnetic polarity reversal of Solar Cycle 24 was relatively slow and took almost $\sim$16-months to switch completely from one cycle to the other. 

In what follows, we discuss aspects of the magnetic field reversal, the relevant GCR observations and very local interstellar spectra, the numerical model with its specific modulation parameters and how they relate to solar activity, followed by corresponding simulations and detailed comparisons with observations for the mentioned period.
 
\section{Solar activity and the Sun's magnetic field reversal} \label{sec:polarity}

Solar activity varies continuously and follows an $\sim$11-year cycle characterised by, amongst others, the numbers and surface area of sunspots. Other solar activity indicators such as the magnitude of the heliospheric magnetic filed (HMF), the tilt of the heliospheric current sheet (HCS), the 10.7 cm radio flux, frequency of solar flares and coronal mass ejections (CMEs), the total solar irradiance etc. are also varying in association with sunspots, see e.g. \citet{2013LRSP...10....1U,2015LRSP...12....4H}.   

\begin{figure}[ht!]
\plotone{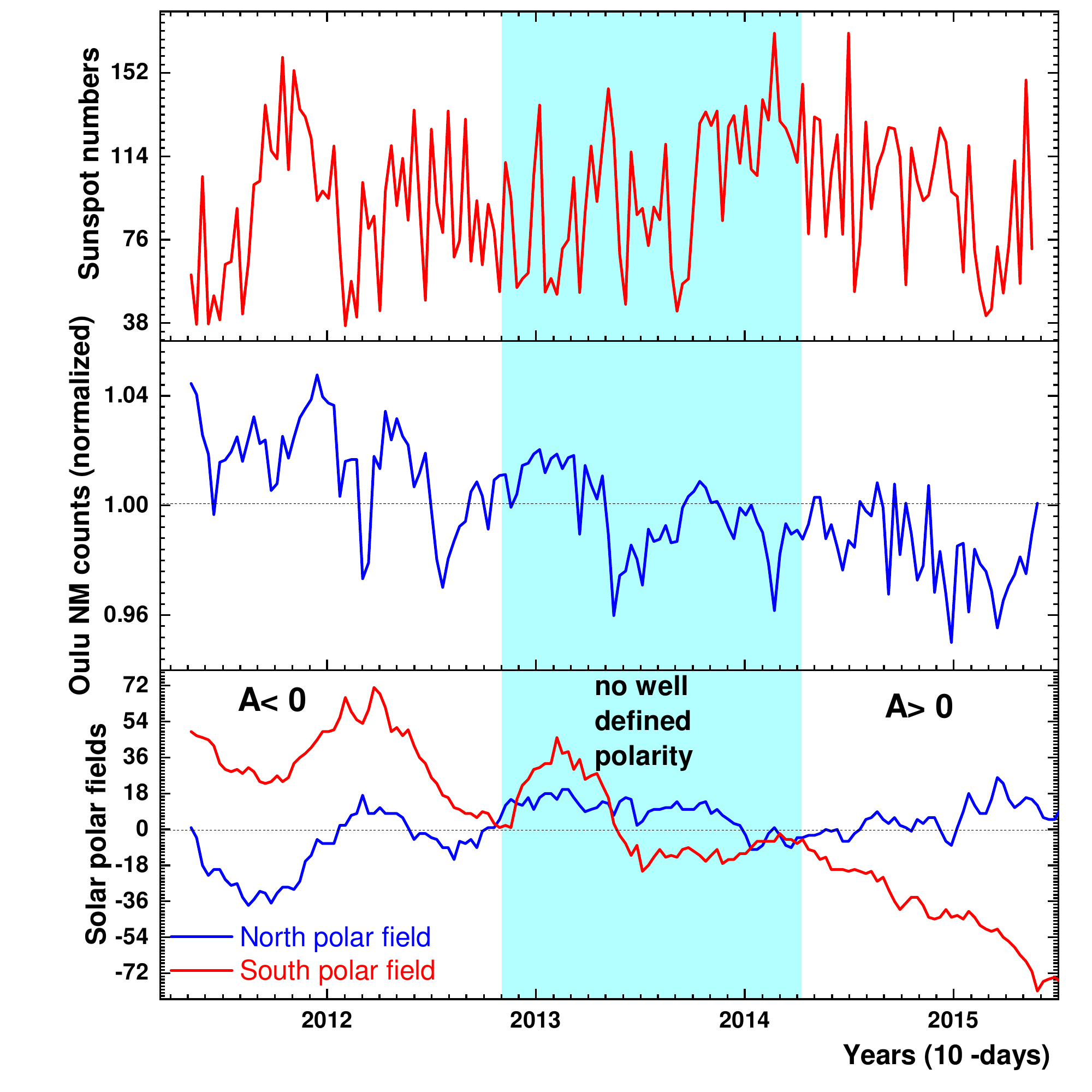}
\caption{Temporal variation of Sun's polar fields of a 10-day resolution, North (blue) and South (red) from Wilcox Solar Observatory records (\url{http://wso.stanford.edu/}) for May 2011 to May 2015 (lower panel). For the same period, counts from the Oulu neutron monitor (\url{https://cosmicrays.oulu.fi/}) also of a 10-day resolution, normalized with respect to the average of May 2011 to May 2015 (middle panel), and the corresponding 10-day-averaged Sunspot Numbers (\url{https://www.sidc.be/silso/home}) (upper panel).
The shaded portion indicates the period of magnetic field polarity reversal when there was no well-defined polarity, here changing from $A < 0$ to $A > 0$; see \citet{2015ApJ...798..114S}. \label{fig1}}
\end{figure}

During solar minimum activity conditions, the magnetic flux concentrations in the solar polar regions are predominantly unipolar, i.e., a dipole-like large-scale solar magnetic field, but as the solar cycle progresses, the magnetic flux from active regions migrates pole ward to replace the old-cycle polar field, and eventually the polar field and the dipole moment reverse sign around the maximum activity phase; see e.g. \citet{1959ApJ...130..364B, 2008ApJ...688.1374T}.
\citet{2015ApJ...798..114S} described the Sun's magnetic field's reversal process during the recent Solar Cycle 24 using the line-of-sight magnetic field measurements from the Helioseismic and Magnetic Imager, and concluded that the field reversal, the so-called polarity reversal, was relatively slow, with the northern and southern polar fields reversed in 2012 November and 2014 March, respectively. 
Until November 2012, the magnetic polarity was negative (denoted as A$<$0); then, the magnetic field lines are directed outward in the Sun’s southern hemisphere and inward in the northern hemisphere.  Since March 2014, the polarity is positive (denoted as A$>$0); the field lines are directed outward in the Sun’s northern hemisphere and inward in the southern hemisphere. The northern and southern polar fields reversed 16-months apart, which is a rather slow rate. This period is shaded in Figure \ref{fig1} and indicates the period of polarity reversal when no well-defined polarity was present.  
The temporal variation of Sun's northern (blue) and southern (red) polar fields of a 10-day resolution observed at Wilcox Solar Observatory (\url{http://wso.stanford.edu/}) from May 2011 to May 2015 is shown in the lower panel of Figure \ref{fig1}. The polarity is quite clear before November 2012 and after March 2014.
Shown in the top panel of Figure \ref{fig1} are Sunspot Numbers here for 10-day averages (\url{https://www.sidc.be/silso/home}) again from May 2011 to May 2015. Although these sunspot numbers fluctuate significantly, it is known that Solar Cycle 24 was 30\% weaker compared to the previous cycle (\url{https://www.sidc.be/silso/news004}); see \citet{2015LRSP...12....4H} for a detailed review about this Solar Cycle. At a solar rotation average scale (Carrington rotations/27-days) the progress and peak of Solar Cycle 24 are clearer in Sunspots Numbers and especially in the tilt angle ($\alpha$) of the HCS, as we show later in Figure \ref{fig4} in the context of what is required for numerical modeling for this period.
Shown in the mid-panel of Figure \ref{fig1}, from May 2011 to May 2015 of a 10-day resolution, is the counting rate of the Oulu neutron monitor (NM), which is located in Oulu, Finland with a geomagnetic cut-off rigidity of $\sim$0.80 GV (geolatitude 65.05$^{\circ}$N, geolongitude 25.47$^{\circ}$E, altitude 15 m asl). The counts are normalized with respect to the May 2011 to May 2015 period average (\url{https://cosmicrays.oulu.fi/}).

When considering long term modulation and observations on a BR resolution scale, it is expected that transient variations in GCR flux caused by interplanetary counter parts of coronal mass ejections (ICMEs; sudden release of plasma along with magnetic field from the solar corona to the heliosphere), e.g., Forbush decreases \citep{1937PhRv...51.1108F} will be smoothed out because they are usually sudden with complete recovery in a few days. See also for example the simulations of Forbush decreases done by \citet{2017ApJ...839...53L}.
\citet{2000SSRv...93...55C} summarized the characteristics of CMEs, explaining how they affect GCRs and the mechanisms involved in producing transient variations, and also discussed the relative role of CMEs in long-term modulation. For recent updates on Forbush decreases associated with ICMEs, see e.g., \citet{2020ApJ...896..133L}, \citet{2021Ap&SS.366...62B, 2021AdSpR..68.4702B}, \citet{2022MNRAS.515.4430M}, and \citet{2021ApJ...920L..43A} for Forbush decrease observations above 2 GV 
from the Dark Matter Particle Explorer (DAMPE) experiment, and \citet{2018ApJ...853...76M} for PAMELA observations. It should also be noted that these transients may merged into larger structures further out in the heliosphere \cite[e.g.,][and references there-in]{1999AdSpR..23..501L}, the effects of which on GCR modulation were simulated and illustrated also recently by \citet{2019ApJ...878....6L}; for simulations of corotating interaction regions, see \citet{2020ApJ...899...90L}.
However, the effects of the occurrence and causes of these transients on GCRs observed at the Earth are not the focus of our current modeling.

\begin{figure}[ht!]
\plotone{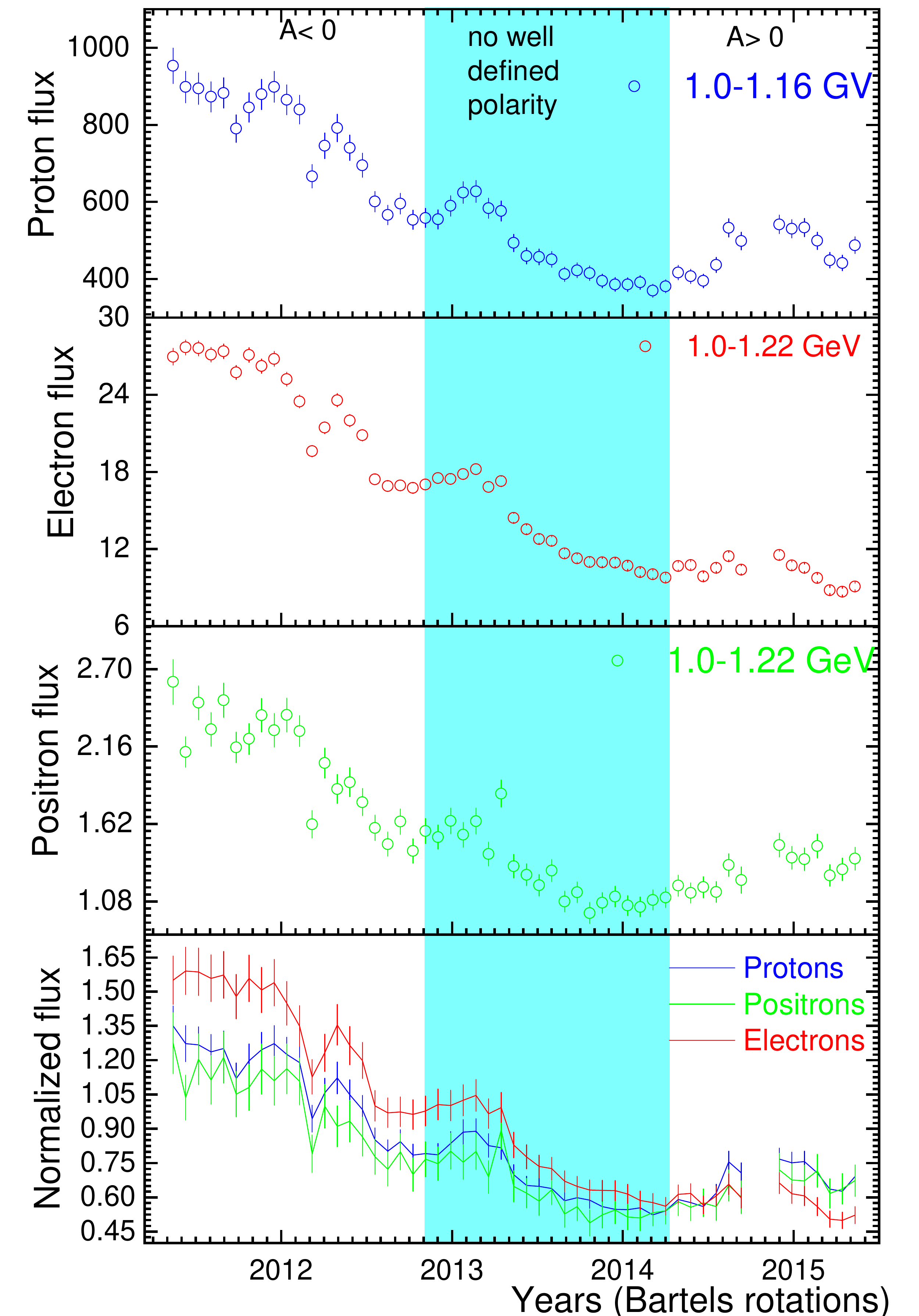}
\caption{Bartels rotation averaged GCR flux from AMS02 for May 2011 to May 2015 (Bartels Rotation number: 2426 - 2480). Top panel: protons for the rigidity range of 1.0 - 1.16 GV as reported by \citet{2018PhRvL.121e1101A}. Second and third panels: electrons and positrons for the kinetic energy range 1.0 - 1.22 GeV as reported by \citet{2018PhRvL.121e1102A}. Bottom panel: All three fluxes for comparison on the same scale, normalized with respect to the average over May 2011 to May 2015. Shaded parts again indicate the polarity reversal phase of Sun's magnetic field, changing from $A<0$ to $A>0$. \label{fig2}}
\end{figure}
\section{Cosmic ray observations} \label{sec:Observations}

The Alpha Magnetic Spectrometer (AMS) is a  large acceptance precision particle detector on the International Space Station, design to measure the sign and value of the charge, the momentum, the rigidity, and the flux of cosmic ray particles directly in space, i.e., electrons, positrons, protons, antiprotons, GCR nuclei and anti-nuclei. \citet{2013PhRvL.110n1102A} reported the positron fraction in the kinetic energy range of 0.50-350 GeV and \citet{2016PhRvL.117i1103A} reported the antiproton flux and the antiproton-to-proton flux ratio in the rigidity range from 1.0-450 GV, over the time period 19 May 2011 to 26 May 2015. The availability of these precise long-term observations with different time resolutions and different rigidity ranges provide incentives for detailed studies of various aspects of heliospheric modulation, in particular drift effects over complete solar cycles. Apart from AMS observations distinguishing between GCRs particles according to their charge and mass, they are available up to very high rigidity \citep{2021PhR...894....1A}. The latter and similar observations allow for comparative studies with NMs, e.g., the geomagnetic cut-off rigidity of the Oulu NM is $\sim$0.80 GV which means it records cosmic particles at 0.80 GV and above \citep{1991ICRC....3..145K}. However, as such, NMs measure integrated intensities of cosmic rays depending on the NM's location in terms of the Earth's magnetosphere and atmosphere, that is, the location above sea level; see e.g., \citet{2015JGRA..120.7172G, 2017JGRA..122.3875U}. These aspects make comparative interpretations rather difficult, as we elaborate on when exhibiting Figures \ref{fig1}, \ref{fig2}, and \ref{fig6}. For space bound observations, interpretation is relatively easier, for example, the lowest fluxes observed for protons and positrons were at the end of the polarity reversal period, between October 2013 and February 2014, whereas for electrons and antiprotons the lowest fluxes were observed later, around March 2015.

Long-duration space based experiments with sophisticated detectors also resolve GCR species and their isotopic composition over a wide range of rigidity with a relatively high time resolution, e.g. PAMELA \citep{2017NCimR..40..473P}, AMS \citep{2021PhR...894....1A}, and DAMPE \citep{2021ApJ...920L..43A}. Resolving the charge of GCRs is most helpful for studies of charge-sign dependant modulation over extended periods of time, which is what we focus on, especially over the HMF polarity reversal phase of Solar Cycle 24. 

\citet{2018PhRvL.121e1101A} reported GCR proton and helium fluxes averaged over Bartels rotations (BR) for May 2011 to May 2017 over the rigidity range 1.0-60.3 GV for protons, and 1.92-60.3 GV for helium. \citet{2018PhRvL.121e1102A} reported BR averaged electron and positron fluxes for May 2011 to May 2017 over the kinetic energy range 1.0-50 GeV. We also make use of the antiproton flux reported by \citet{2016PhRvL.117i1103A} along with the reported antiproton flux for 60 MeV-350 GeV obtained from June 2006 to January 2010 by the PAMELA experiment; see \citet{2013JETPL..96..621A, 2017NCimR..40..473P} for detailed descriptions of the PAMELA detector and mission.

In this study, we specifically utilize BRs averages for proton, electron and positrons observations from May 2011 to May 2015 (BRs 2426 to 2480) for modeling, but not available for anti-protons. This is then combined with simulated antiproton spectra for each BR for the same period. This selection covers the solar magnetic polarity reversal phase of Solar Cycle 24 (November 2012-March 2014).
These data sets are used to validate our modeling parameters and assumptions of local interstellar spectra, especially for anti-protons, as will be shown and discuss in the next section. 

In Figure \ref{fig2} the AMS fluxes, averaged over BRs, for protons (1.0-1.16 GV), electrons (1.0-1.22 GeV) and positrons (1.0-1.22 GeV) are plotted separately in the top three panels, and together in the bottom panel as normalized fluxes, from 2011 May to 2015 May. It follows from the latter comparison that these particles exhibit similar patterns with time but electrons had a relative higher flux from 2011 May during the A$<$0 phase to end of polarity reversal period, but in the A$>$0 phase the electron flux has become relatively lower. We focus on this aspect, considering it as indication of the change in the drift patterns for these oppositely charged GCRs from before to after the polarity reversal.
In this context, \citet{2021ApJ...909..215A} showed how the positron to electron flux ratio varied from June 2006 to December 2015, including the polarity reversal phase, and how particle drift had changed over this polarity reversal, using the same 3D numerical modulation model as described below. They compared the modeling results to the PAMELA positron to electron flux ratio reported by \citet{2016PhRvL.116x1105A} and AMS positron and electron fluxes reported by \citet{2018PhRvL.121e1102A}. In a similar way, \citet{2021Potgieter} illustrated from a modeling perspective how the antiproton to proton flux ratio varied over the Sun's recent magnetic polarity reversal.

\section{Very Local Interstellar Spectra of Galactic Cosmic rays} \label{sec:LIS}

In the numerical modeling of the modulation of GCRs it is required to specify a galactic spectrum for each and every species that is studied. More specifically, for heliospheric modulation studies a very local interstellar spectrum (vLIS) needs to be specified at the simulated outer heliospheric boundary as an initial condition, i.e., an input spectrum that is to be modulated from this outer boundary up to the Earth. The heliopause (HP) is considered to be this outer boundary and is assumed to be located at 122 au from the Sun based on the reports from the Voyager 1 and Voyager 2 space missions \citep{2013Sci...341..150S, 2019NatAs...3.1013S}. Because the total modulation is dependent on the global size of the modulation region, confirming the existence of the HP and settling on 120-122 au as outer boundary was a major step forward in numerical modeling; see also \citet{2017AdSpR..60..848P,2019AdSpR..64.2459B}. 

\begin{figure*}
\gridline{\fig{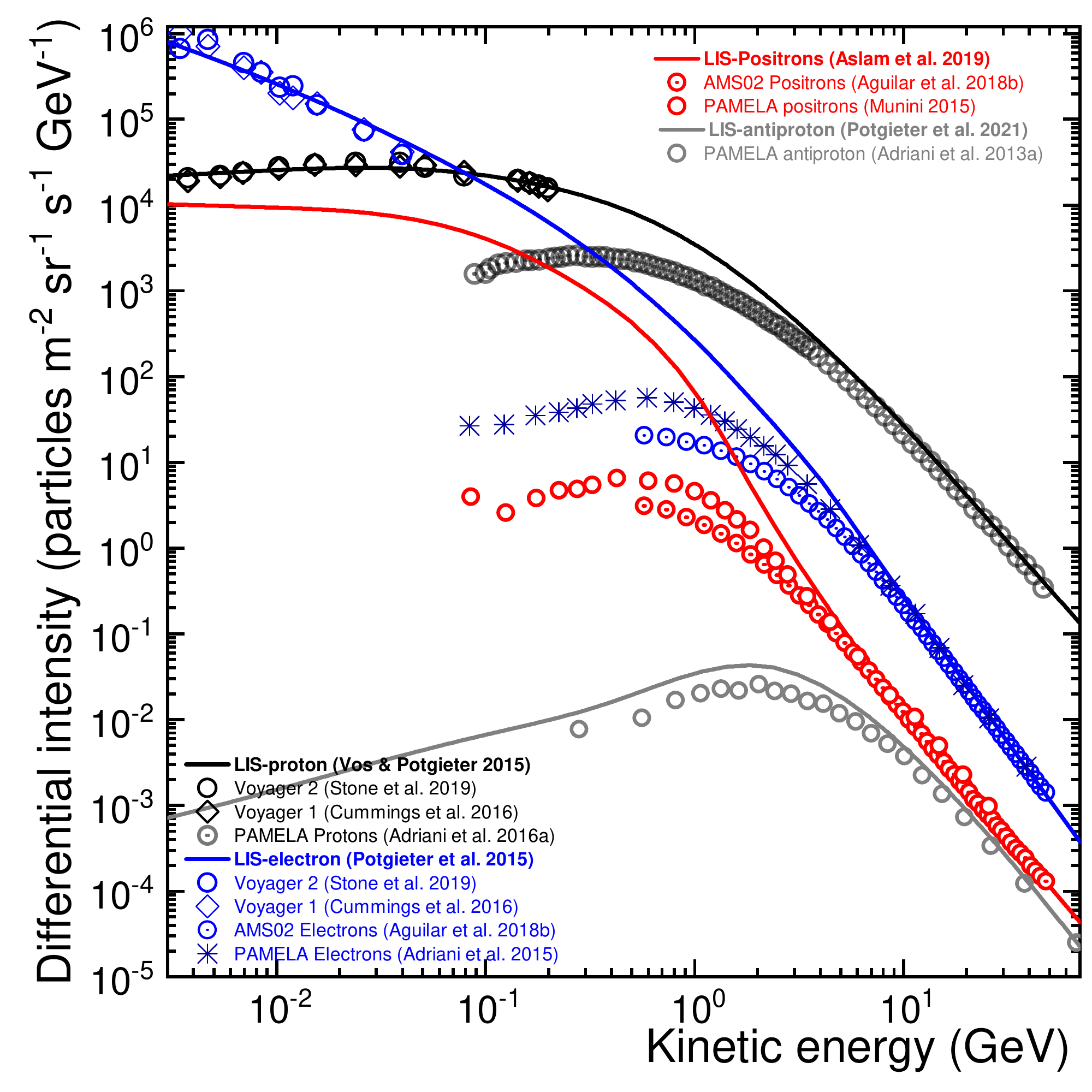}{0.5\textwidth}{(a)}
       \fig{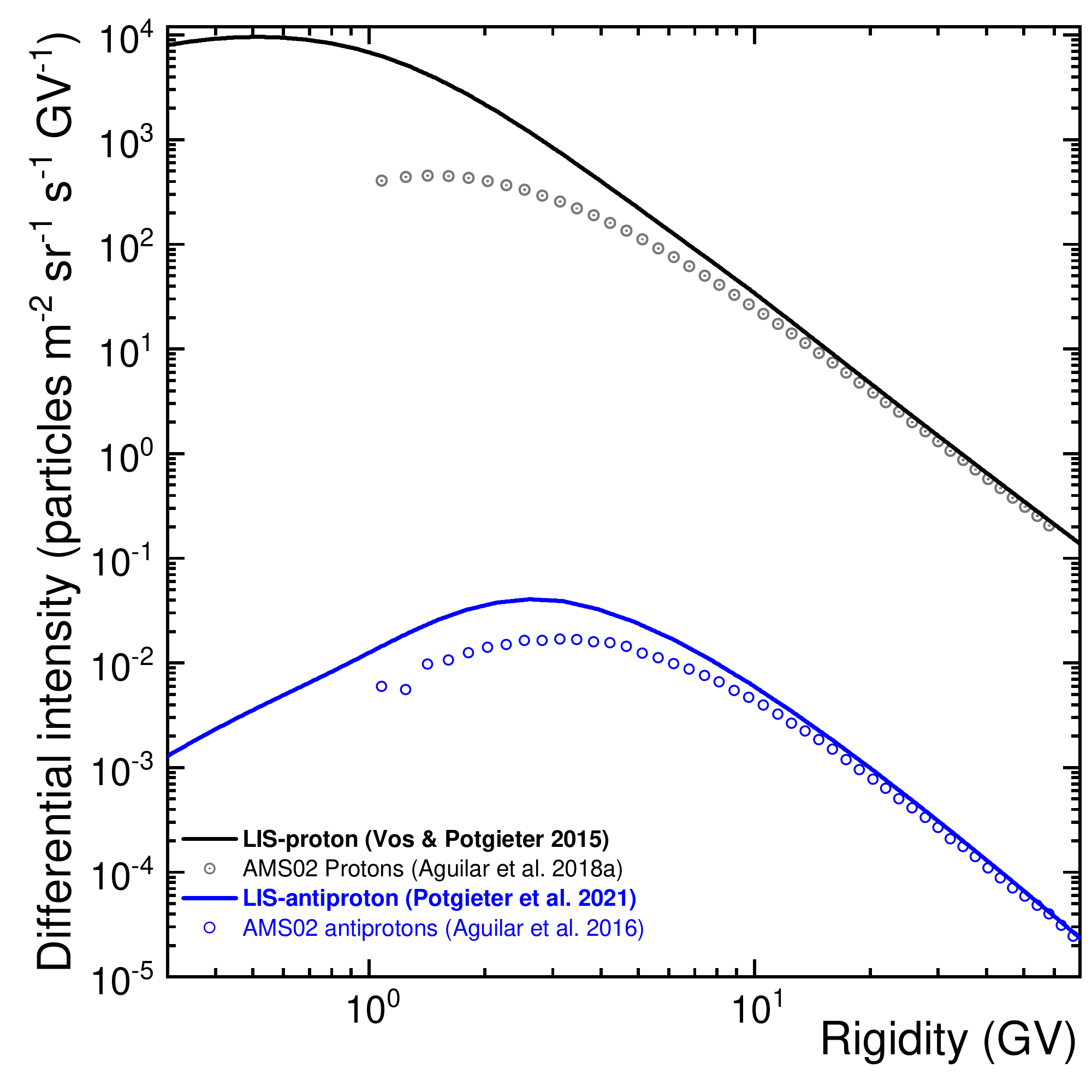}{0.5\textwidth}{(b)}}
\caption{Very local interstellar spectra (vLIS's), shown as a differential intensity per GeV (left panel) and as a differential intensity per GV (right panel), as a function of kinetic energy and rigidity, respectively. First, for GCR protons, shown in both panels, from \citet{2015ApJ...815..119V}; for anti-protons, shown in both panels, from \citet{2021Potgieter}; for GCR electrons, shown in left panel, from \citet{2015ApJ...810..141P}; for positrons, shown in left panel, from \citet{2019ApJ...873...70A}.
These vLIS's are validated by utilizing PAMELA observations \citep{2013JETPL..96..621A, 2013PhRvL.111h1102A, 2015ApJ...810..142A, 2016ApJ...818...68A, Munini2015} and AMS observations at the Earth \citep{2018PhRvL.121e1101A, 2018PhRvL.121e1102A} at high kinetic energy/rigidity as indicated, and Voyager 1 and 2 observations \citep{2016ApJ...831...18C, 2019NatAs...3.1013S} at the HP at lower energies, as indicated.
\label{fig3}}
\end{figure*}

The vLIS's used for this numerical study are shown in Figure \ref{fig3} as a differential intensity per GeV (left panel) and as a differential intensity per GV (right panel), as a function of kinetic energy and rigidity, respectively. First, for GCR protons, shown in both panels, from \citet{2015ApJ...815..119V}; for anti-protons, shown in both panels, from \citet{2021Potgieter}; for GCR electrons, shown in left panel, from \citet{2015ApJ...810..141P}; for positrons, shown in left panel, from \citet{2019ApJ...873...70A}. These vLIS's are validated by utilizing PAMELA observations \citep{2013JETPL..96..621A, 2013PhRvL.111h1102A, 2015ApJ...810..142A, 2016ApJ...818...68A, Munini2015} and AMS observations at the Earth \citep{2018PhRvL.121e1101A, 2018PhRvL.121e1102A} at high kinetic energy/rigidity as indicated, and Voyager 1 and 2 observations \citep{2016ApJ...831...18C, 2019NatAs...3.1013S} at the HP at lower energies, as indicated. For various vLIS's, the differences in how they are determined, and their importance for numerical modeling studies, see \citet{2014BrJPh..44..581P,2014ApJ...794..166B,2016ApJ...831...18C,2016ApJ...829....8C,2019ApJ...878...59B,2021PAN....84.1121B,2020ApJ...889..167B,2022AdSpR..69.2330N} and for recent overviews, see \citet{,2021Potgieter} and \citet{2022SSRv..218...42R}.

The shape and values of a vLIS in terms of rigidity/kinetic energy is an important factor in computing the total modulation from the HP to the Earth for a given GCR species. Inspection of Figure \ref{fig3}(b), for example, gives the total modulation for protons at 1 GV to be a factor of $\sim{17}$, whereas for antiprotons it is only $\sim{2}$. This is a fact in modeling despite that the set of modulation parameters used for modulation studies are identical for a given period (solar cycle phase) for protons, antiprotons, all GCR nuclei, and for electrons and positrons. In this context, see also \citet{2014AdSpR..53.1015S,2016SoPh..291.2181V,2022AdSpR..69.2330N}.

\section{Numerical modulation model} \label{sec:model}

We use a sophisticated and comprehensive three-dimensional (3D) steady-state model, which is based on the numerical solution of Parker's transport equation \citep{1965P&SS...13....9P}, to compute the differential intensity of GCRs at different radial distances from Earth (1 au) up to heliopause (122 au). The basic transport equation (TPE) follows from the equation of charged particles motion in a large and small
scales fluctuating magnetic fields, and averages over the pitch and phase angles of propagation particles, with the assumption that GCRs are approximately isotropic. This heliospheric TPE is described as:  
\begin{equation}
\frac{\partial f}{\partial t} = - \vec{V}_{sw} \cdot \nabla \it{f} - \langle \vec {v}_{D} \rangle \cdot \nabla \it{f} + \nabla \cdot (\bf{K}_{s} \cdot \nabla \it{f}) + \frac {1}{3} (\nabla \cdot \vec{V}_{sw}) \frac {\partial f} {\partial  ln  p} \label{Eq1}
\end{equation}
where $f (\vec {r}, p, t)$ is the omnidirectional GCR distribution function, $\it{p}$ is momentum, $\it{t}$ is time, and $\vec {r}$ is the vector position in 3D with the three coordinates $\it{r}$, $\theta$, and $\phi$ specified in a heliocentric spherical coordinate system where the equatorial plane is at a polar angle of $\theta = 90^{\circ}$. If $\it{P}$ is particle rigidity, the intensity is $\it{P}^{2}f$ ($\it{p}^{2}f$). The terms shown on the right-hand side of this equation represent: (1) The outward convection caused by the expanding solar wind with velocity ($\vec {V}_{sw}$), (2) the averaged particle drift velocity $\langle \vec {v}_{D} \rangle$ (pitch angle averaged guiding center drift velocity) i.e. particle drift caused by the HMF's gradients, curvatures and the HCS, which is:
\begin{equation}
\langle \vec {v}_{D} \rangle = \nabla \times K_{D} \frac{\vec B}{B}
\label{Eq2}
\end{equation}
where $K_{D}$ is the generalized drift coefficient, and $\vec{B}$ is the HMF vector with magnitude $\it{B}$. 
(3) Spatial diffusion caused by scattering of GCRs, where $\bf{K}_{s}$ is the symmetry diffusion tensor, and (4) adiabatic energy changes (losses due to the expansion of the solar wind), which depends on the sign of the divergence of $\vec {V}_{sw}$. If $\nabla \cdot \vec{V}_{sw}>0 $, adiabatic energy loss occurs as is the case in most of the heliosphere, except inside the heliosheath where we assume that $\nabla \cdot \vec{V}_{sw} = 0$; see also \citet{2006ApJ...640.1119L}.
As mentioned above, all four known physical processes play an important role in modulating GCRs, but their relative contributions vary according to the level of solar activity, including the HMF polarity reversal. 
\citet{2014SoPh..289..391P,2015ApJ...810..141P}, \citet{2019ApJ...873...70A} and \citet{2022AdSpR..69.2330N} gave a detailed description on the 3D numerical modulation model. This steady-state model is not suitable to study GCR events shorter than one solar rotation period as explained by \citet{2021ApJ...909..215A}.
The functional forms of the three diffusion coefficients and the drift coefficient as used in the  model to illustrate the underlying physics, are given and discussed in what follows.
Our results on how these three diffusion coefficients and the drift coefficient change with time (solar activity) are given and illustrated at the end of this paper.

\subsection{Model Specifics: Particle Drift and Diffusion  \label{subsec:Drift}}

The expression for the generalized drift coefficient $K_{D}$ can be rewritten as
\begin{equation} 
K_{D} = \frac {\beta P} {3B_{m}} f_{D} = \frac {\beta P} {3B_{m}} \Bigg[ \frac {(\omega \tau)^{2}}{1+(\omega \tau)^{2}}\Bigg] \label{Eq3}
\end{equation}
where, $B_{m}$ is the magnitude of the Smith-Bieber modified \citep{1991ApJ...370..435S} Parker-type HMF, 
$\beta$ = $v/ c$ is the ratio of particle speed to the speed of light, and $\omega$ is the particle gyro-frequency with $\tau$ the average time between the scattering of GCR particles in the turbulent HMF.
The term $f_{D}$ = $\Bigg[ \frac {(\omega \tau)^{2}}{1+(\omega \tau)^{2}}\Bigg]$ is called the drift reduction factor and is determined by how diffusive scattering is described; if $f_{D}$ = 0, then $K_{D}$ and therefore the drift velocity $\langle v_{D} \rangle$ becomes zero so that drift effects will vanish from the modulation model to produce non-drift solutions; if $f_{D}$ = 1, drift is at a maximum, so that $K_{D}$ then has the weak scattering value because $\omega \tau$ becomes much larger than 1.0, in the order of 10; for detailed discussions, see e.g., \citet{2015AdSpR..56.1525N}, and references there-in. Theoretically, this gives the largest possible value for $K_{D}$, for a given particle rigidity and $B$ value, as found for the unmodified Parker HMF; illustrations of the difference between these HMF models were given by \citet{2016AdSpR..57.1965R}. If this is the case, drift effects in GCR modulation are predicted to be very large and dominant as found in original numerical models by \citet[][where the $A>0$ and $A<0$ computed spectra were the opposite of what was observed]{1979ApJ...234..384J}, \citet{1983ApJ...265..573K} and \citet{1985ApJ...294..425P}, to mention only a few. 
This function $f_{D}$ can be used to adjust the rigidity dependence of $K_{D}$, which is the most effective direct way of suppressing drift effects at low rigidity as required by $\it{Ulysses}$ observations of latitudinal gradients; see reviews by \citet{2001SSRv...97..309H}; \citet{2006SSRv..127..117H}; \citet{2013SSRv..176..265H}. As such, Equation (\ref {Eq3}) gets a more practical form \citep[e.g.][]{2015AdSpR..56.1525N,2018ApJ...859..107M} and has been widely used in numerical drift models:

\begin{equation}
K_{D} = K_{A0}  \frac {\beta P} {3B_{m}} \frac {(P/P_{A0})^{2}}{1+(P/P_{A0})^{2}} \label{Eq4}
\end{equation}           
Here, $K_{A0}$ is dimensionless, and may be ranging from 0.0 to 1.0 as will be illustrated in a later section.

The expression for the diffusion coefficient parallel to the average background HMF, as used in this modeling effort, is given by:

\begin{equation}
K_{\parallel} = (K_{\parallel})_{0} \beta \Bigg(\frac {B_{0}}{B_m}\Bigg) \Bigg(\frac {P}{P_{0}}\Bigg)^{c_{1}} \left[ \frac {\Bigg(\frac {P}{P_{0}}\Bigg)^{c_{3}} + \Bigg(\frac {P_{k}}{P_{0}}\Bigg)^{c_{3}}}{ 1+ \Bigg(\frac {P_{k}}{P_{0}}\Bigg)^{c_{3}}} \right]^{\frac {c_{2 \parallel} - c_{1}}{c_{3}}}
\label{Eq5}
\end{equation}
where $(K_{\parallel})_{0}$ is a scaling constant in units of $10^{22}$ cm$^{2}$s$^{-1}$, with the rest of the equation written to be dimensionless with $P_{0} = 1.0$ GV, and $B_{0}$ = 1.0 nT (in order to preserve the units in cm$^{2}$ s$^{-1}$). Here $c_{1}$ is a power index that may change with time; $c_{2 \parallel}$ and $c_{2 \perp}$ together with $c_{1}$ determine the slope of the rigidity dependence, respectively, above and below a rigidity with the value $P_{k}$ which may also change with time; $c_{3}$ determines the smoothness of the transition. The rigidity dependence of $K_{\parallel}$ is thus a combination of two power laws; $P_{k}$ determines the rigidity where the transition in the power laws occur, and the value of $c_{1}$ determines the slope of the power law at rigidity below $P_{k}$. How the rigidity dependence changes with time in particular  will be shown later. 
Note that the relation between diffusion coefficients ($K$) and their corresponding mean free paths (MFPs; $\lambda$) in general is: $K$ = $\lambda$($v/3$), where $v$ is the speed of the particles. 

Perpendicular diffusion in the radial direction is assumed to scale spatially similar to Equation (\ref {Eq5}) but with a different rigidity dependence at higher rigidity:

\begin{equation}
K_{\perp r} = 0.02 (K_{\parallel})_{0}  \beta  \Bigg(\frac {B_{0}}{B_{m}}\Bigg) \Bigg(\frac {P}{P_{0}}\Bigg)^{c_{1}} \left[ \frac {\Bigg(\frac {P}{P_{0}}\Bigg)^{c_{3}} + \Bigg(\frac {P_{k}}{P_{0}}\Bigg)^{c_{3}}} { 1+ \Bigg(\frac {P_{k}}{P_{0}}\Bigg)^{c_{3}}} \right]^{\frac {c_{2 \perp r}- c_{1}}{c_{3}}}.
\label{Eq6}
\end{equation}
Using $K_{\perp r}$ = 0.02 $K_{\parallel}$ is a widely used and reasonable assumption, e.g. \citet{1999ApJ...520..204G}. In the case of electrons and positrons, the value of c$_{1}$ = 0.0 which accounts for a significant difference between these GCR particles and protons and antiprotons at lower rigidity.  See \citet{2016AdSpR..58..453N} and \citet{2019ApJ...873...70A,2021ApJ...909..215A} for illustrations of the modulation effect on electrons when changing these parameters. 

The polar perpendicular diffusion ($K_{\perp \theta}$), on the other hand is more complicated, with consensus that $K_{\perp \theta}$ $>$ $K_{\perp r}$ away from the equatorial regions as discussed and motivated by \citet{2000JGR...10518295P} and \citet{2014SoPh..289..391P}. 
In  this study, perpendicular diffusion in the polar direction is assumed to scale spatially similar to Equation (\ref {Eq6}) but with a different rigidity dependence at higher rigidity:

\begin{equation}
K_{\perp \theta} = 0.02 f_{\perp \theta} (K_{\parallel})_{0}  \beta  \Bigg(\frac {B_{0}}{B_m}\Bigg) \Bigg(\frac {P}{P_{0}}\Bigg)^{c_{1}} \left[ \frac {\Bigg(\frac {P}{P_{0}}\Bigg)^{c_{3}} + \Bigg(\frac {P_{k}}{P_{0}}\Bigg)^{c_{3}}} { 1+ \Bigg(\frac {P_{k}}{P_{0}}\Bigg)^{c_{3}}} \right]^{\frac {c_{2 \perp \theta}- c_{1}}{c_{3}}},
\label{Eq7}
\end{equation}
with
\begin{equation}
f_{\perp \theta} = A^{+} \mp A^{-} tanh[8(\theta_{A} - 90^{\circ}) \pm \theta_{F}].
\label{Eq8}
\end{equation}
Here, A$^{\pm} = (d_{\perp \theta} \pm 1)/2 $, $\theta_{F}$ = 35$^{\circ}$, $\theta_{A}$ = $\theta$ for $\theta \leq 90^{\circ}$ but  $\theta_{A}$ = 180$^{\circ}$ - $\theta$ with $\theta \geq$  90$^{\circ}$ and $d_{\perp \theta}$ = 6.0 for protons and antiprotons, 9.0 to 3.0 for electrons and positrons. 
This means that $K_{\perp \theta}$ has a different latitudinal dependence than the other diffusion coefficients which can be enhanced towards the heliospheric poles by a factor $d_{\perp \theta}$ with respect to the value of $K_{\perp r}$ in the equatorial region of the heliosphere. We assume $d_{\perp \theta}$ = 6 for the period before the solar polarity reversal, then gradually change it to 9.0 in the case of electrons and positrons, but keep $d_{\perp \theta}$ = 6 for the whole period in the case of protons and antiprotons. For a detailed theoretical discussion of these aspects of the diffusion tensor and theory, see the extensive reviews by \citet{2009ASSL..362.....S,2020SSRv..216...23S}. 
For a motivation and applications of this particular modeling approach, see \citet{2000JGR...10518295P}, \citet{2003AnGeo..21.1359F}, \citet{2012AdSpR..49.1660N}, \citet{2016AdSpR..58..453N}, to mention only a few.

\subsection{Model Specifics: Solar activity dependence\label{subsec:intrinsic}}

\begin{figure}[ht!]
\plotone{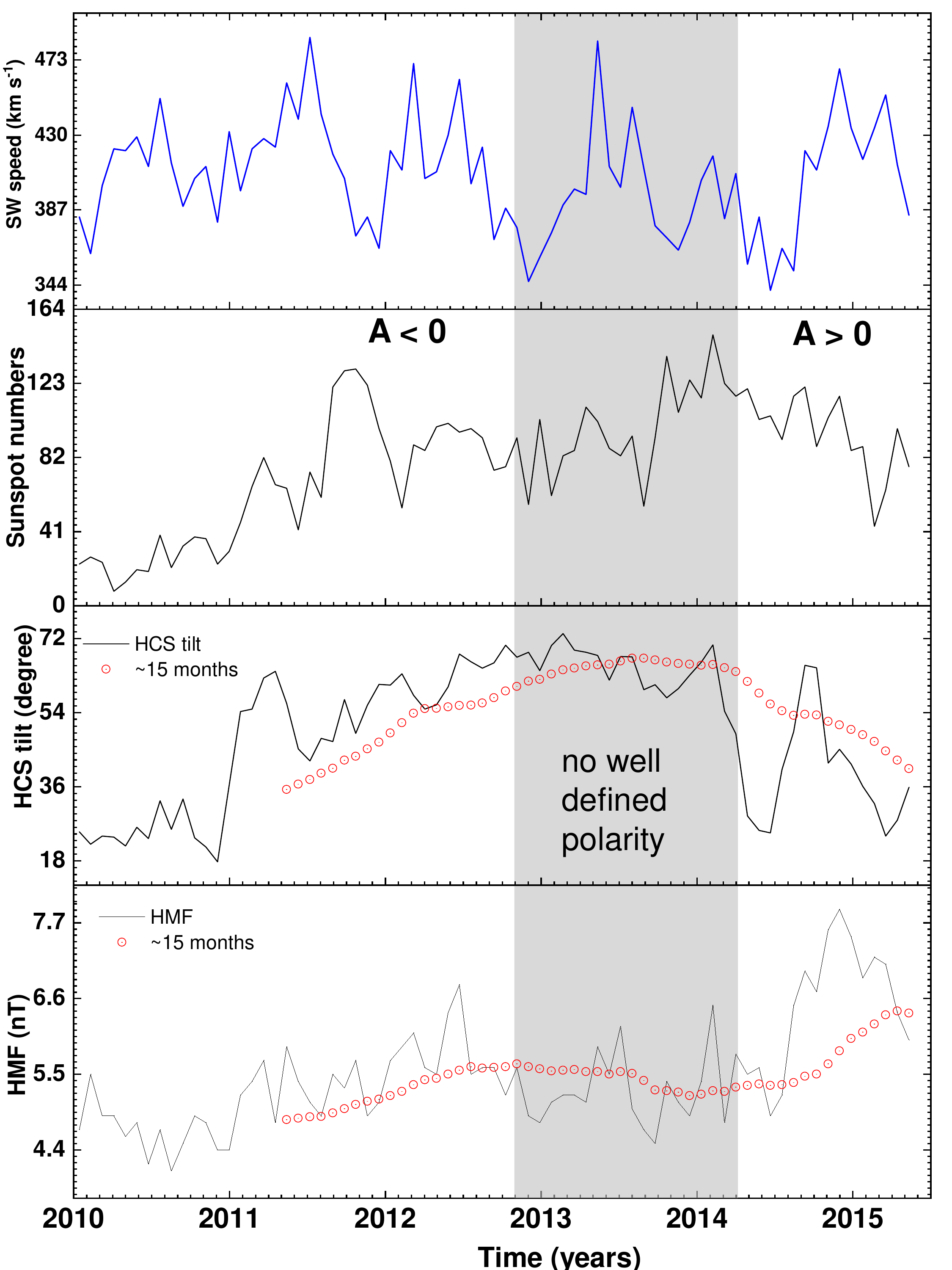}
\caption{Lower panel: Magnitude of the HMF observed at the Earth averaged over Bartels rotations (BR), adopted from \url{http://omniweb.gsfc.nasa.gov} for January 14, 2010 to May 12, 2015 (BRs 2408 to 2480) shown as solid line, together with $\sim$15-months (17-BRs) running averages, shown as red circles, used in the model as intrinsic parameter.
Second lower panel: Tilt angle ($\alpha$) of the Heliospheric Current sheet (HCS) as Carrington rotation (CR) averages adopted from \url{http://wso.stanford.edu/} for January 30, 2010 to May 21, 2015 (CRs 2093 to 2164) shown as solid line with red circles the $\sim$15-months (17 CRs) running averages used in the model as intrinsic parameter. 
Second upper pane): Observed BR averaged International Sunspot Number adopted from \url{http://omniweb.gsfc.nasa.gov} for January 14, 2010 to May 12, 2015 (BRs 2408 to 2480). 
Top panel: Similar BR averaged solar wind speed observations adopted from \url{http://omniweb.gsfc.nasa.gov} for January 14, 2010 to May 12, 2015.
Shaded parts indicate the polarity reversal phase of Sun's magnetic field, when changing from $A<0$ to $A>0$.
\label{fig4}}
\end{figure}

Observational data, as proxies for solar activity, are used as input to the model and shown together in Figure \ref{fig4}. 
First, shown as a solid line, in the bottom  panel, is the magnitude of the HMF observed at the Earth and averaged over BRs taken from \url{http://omniweb.gsfc.nasa.gov}, from January 14, 2010 to May 12, 2015 (BRs 2408 to 2480) together with $\sim$15-months (17-BRs) running averages, shown as red circles, and used in the model as intrinsic modulation parameter; $B_m$ in the equations given above.
In the second lower panel, the 'radial' tilt angle ($\alpha$) of the HCS is shown as a solid line for Carrington rotation (CR) averages adopted from \url{http://wso.stanford.edu/} for January 30, 2010 to May 21, 2015 (CRs 2093 to 2164), shown together with red circles as $\sim$15-months (17-CRs) running averages used in the model as second proxy for solar activity. In the second upper panel, the observed BR averaged International Sunspot Number are shown from January 14, 2010 to May 12, 2015 (BRs 2408 to 2480), adopted from \url{http://omniweb.gsfc.nasa.gov} (see \url{https://www.sidc.be/silso/home} for the original source) as additional indicator of solar activity. Shown in the top panel is the observed BR averaged solar wind speed (interplanetary plasma speed) adopted from \url{http://omniweb.gsfc.nasa.gov} for January 14, 2010 to May 12, 2015.
As in previous figures, the shaded parts indicate the polarity reversal phase of Sun's magnetic field when  changing from $A<0$ to $A>0$.

In order to simulate BR 2426 (May-June, 2011), the value of $\alpha$ and the averaged HMF magnitude at Earth are determined as 35.38$^{\circ}$ and 4.84 nT, respectively. As the Solar Cycle progresses towards solar maximum, $\alpha$ increases gradually up to 67.25$^{\circ}$ (BR number 2456; July-August, 2013), then gradually decreases and reaches 40.39$^{\circ}$ by April-May, 2015 (BR number 2480), see Figure \ref{fig4}. At first, the magnetic field shows the same trend as $\alpha$, i.e., a gradual increase but only up to the start of the polarity reversal (5.65 nT; BR number 2446, October-November 2012), then it shows a slight decreasing trend (but with almost the same values; changing from 5.65 nT to 5.31 nT) for the whole polarity reversal period. After the reversal, the magnetic field shows a systematic increase to reach 6.42 nT by March-April 2015 (BR number 2479). This behaviour is unlike $\alpha$, which exhibits a rather large drop in mid-2014, similar to what can be expected during solar minimum, only to increase rapidly again to a large value before gradually decreasing. This magnetic field behaviour was probably caused by the increase in the frequency of large interplanetary coronal mass ejections, see \url{https://izw1.caltech.edu/ACE/ASC/DATA/level3/icmetable2.htm}, and needs further investigation beyond the scope of our modulation modeling.

Based on the Voyager 2 measurements, \citet{2011ApJ...734L..21R} reported the dynamic nature of the termination shock, i.e. moving inward and outward in response to changes in the dynamic pressure of the solar wind; this changing width of the inner heliosheath affects the modulated GCR intensity at the Earth. Based on these reports, and in addition to changing $\alpha$ and the HMF magnitude, we also incorporate this dynamic nature of the termination shock (TS) in the model; see \citet{2005AdSpR..35.2084L}, \citet{2014SoPh..289.2207M}, \citet{2016SoPh..291.2181V}.
In the model, the position of the TS is changed gradually from 83 to 88 au from May 2011 to April-May 2014, then gradually decreased to 86 au by April-May 2015.

During solar minimum activity conditions such as the deep minimum between Solar Cycle 23 and 24, the global latitudinal dependence of the solar wind velocity is assumed to change from 430 km s$^{-1}$ in the equatorial plane to 750 km s$^{-1}$ in the polar regions, in accordance with Ulysses observations \citep{2002GeoRL..29.1290M, 2006SSRv..127..117H}. In the model, as the Solar Cycle progresses towards  maximum, this latitudinal dependence of the solar wind velocity gradually decreases in the polar region to become to 450 km s$^{-1}$ when $\alpha$ is 65$^{\circ}$ or higher.


\section{Modeling results and comparison with Observations} \label{sec:results}

The main objectives of this study, to recap, are to study with numerical modeling the unfolding of GCR drift effects from an A$<$0 to A$>$0 polarity phase through the transition period of $\sim$16-months of uncertain magnetic polarity.
We aim to improve our  understand of how the major physical modulation processes had changed over this peculiar HMF polarity reversal time for different GCR species, and if any modulation process could be considered to dominate over this period, as well as understanding to what extent particle drift had changed in order to resemble the observed behaviour of oppositely charged particles.
As such, we simulate and then reproduce every BR averaged AMS proton, electron and positron fluxes, from May 2011 to May 2015 (BRs 2426-2480), and subsequently simulate the antiproton flux for the same BRs, applying the model over the rigidity range 10 MV to 70 GV. 

Because our 3D modeling is done in a steady-state, it is to be expected that the simulated intensity for every BR at various rigidity for the GCR particles studied here are not reproducing the observed BR fluxes always equally well. This is remedied, to a large degree, by working with the averages for the input parameters as depicted in Figure \ref{fig4}. It should be kept in mind that the modulation of GCRs observed at the Earth is determined by modulation conditions between the Earth and the HP.

\subsection{Modulation from May 2011 to May 2015} \label{sec:Modulation 2011 to 2015}

After simulating and reproducing the AMS proton and electron fluxes of each BR from 2426 to 2480, we applied the same parameter set obtained for electrons to reproduce the corresponding positron flux, and the parameter set obtained for protons to simulate the antiproton flux, using the appropriate vLIS's, and changing the magnetic field polarity, apart from the input modulation parameters described above. Comparison is made between modeling results and corresponding observations. 

\begin{figure*}
\gridline{\fig{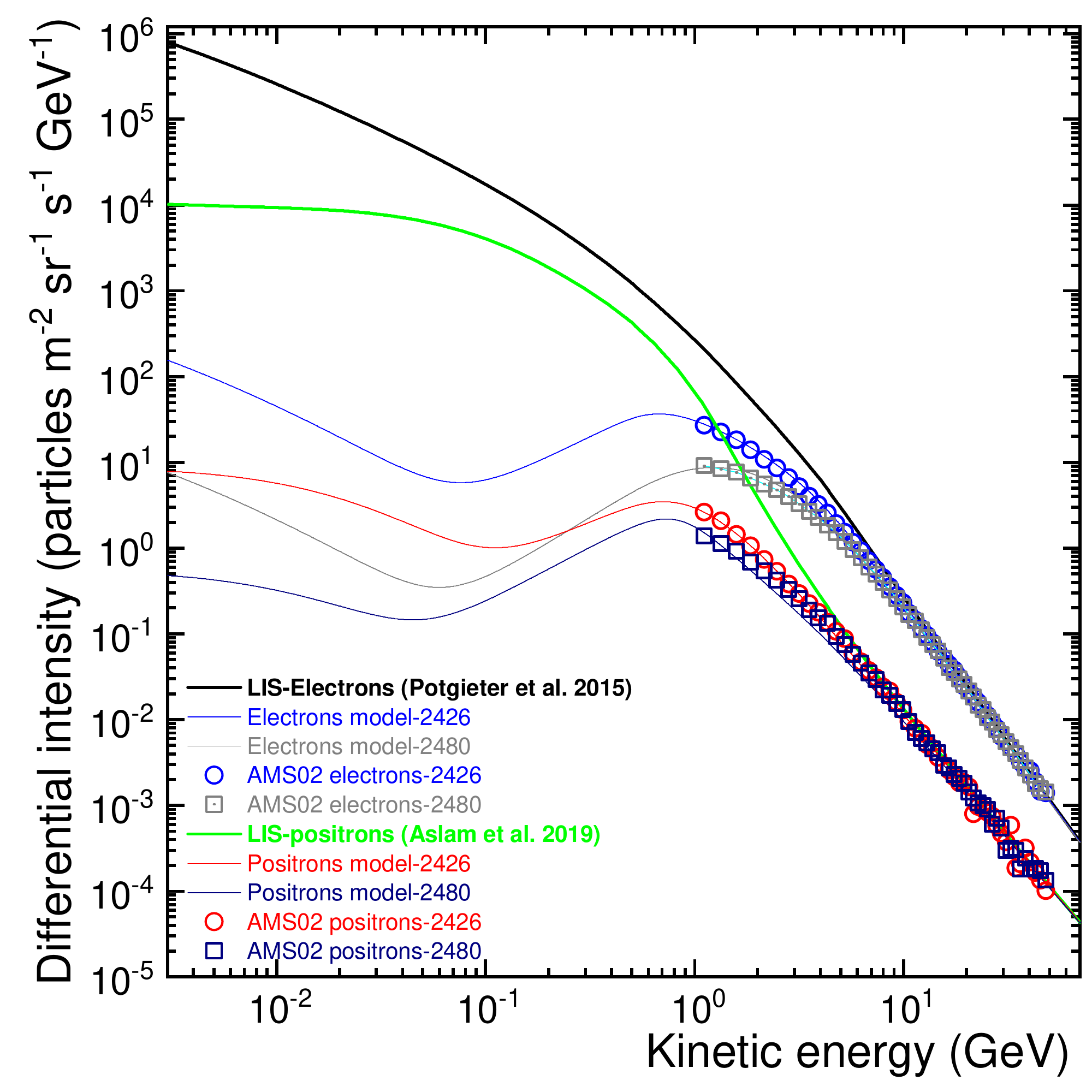}{0.5\textwidth}{(a)}
             \fig{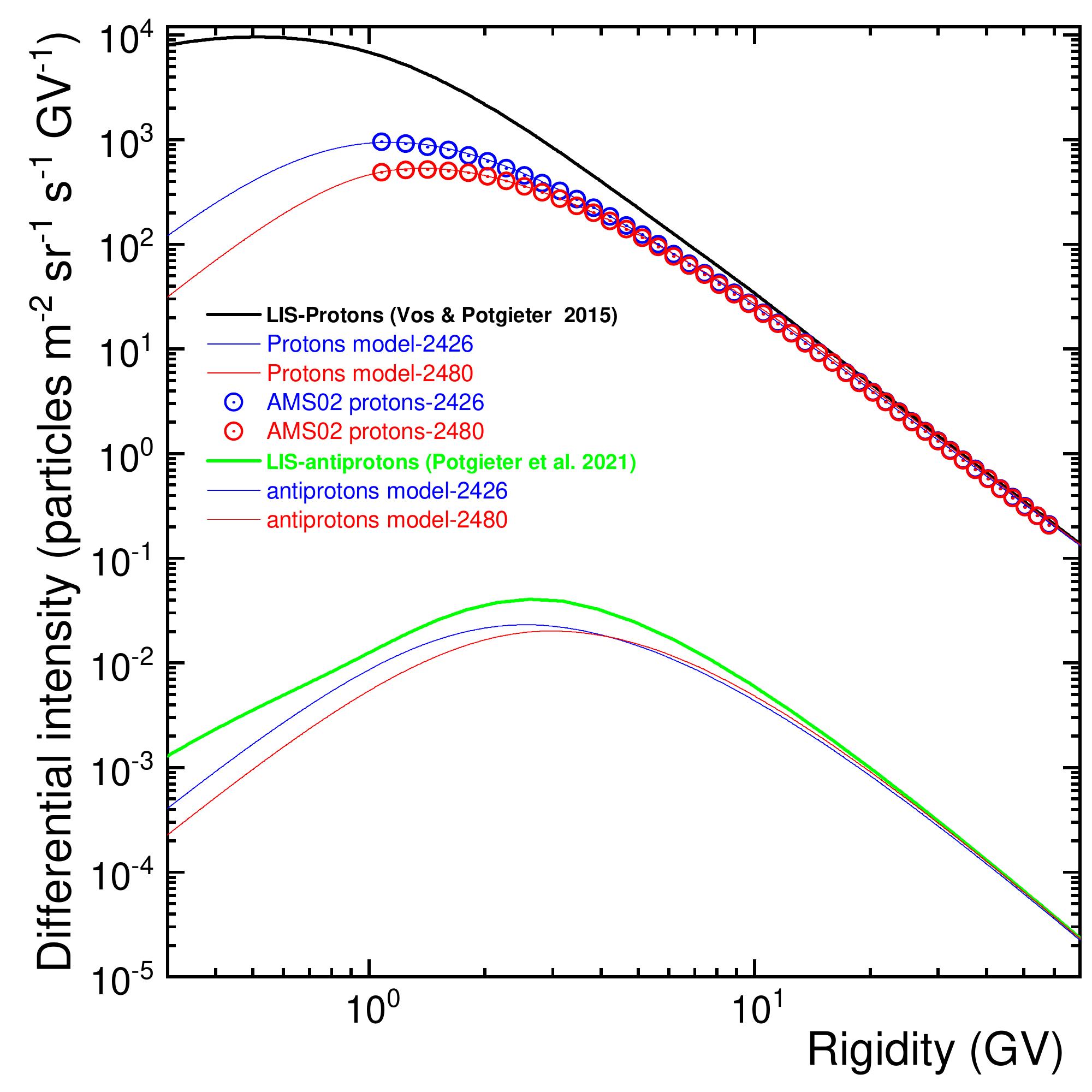}{0.5\textwidth}{(b)}
                    }
\caption{
Panel (a): Simulated spectra for electrons and positrons, BR 2426 (blue \& red lines for electrons \& positrons, respectively) and BR 2480 (gray \& navy lines for electrons \& positrons, respectively), shown at the Earth and compared with AMS observations \citep{2018PhRvL.121e1102A} of the same BRs, blue \& red circles (BR 2426) and gray \& navy squares (BR 2480). The two upper lines, black for electrons and green for positrons, indicate the corresponding vLIS's at 122 au. 
Panel (b): Simulated spectra for protons and antiprotons, BR 2426 (blue solid lines) and BR 2480 (red solid lines), at the Earth and compared with AMS proton observations \citep{2018PhRvL.121e1101A} of the same BRs, blue circles (BR 2426) and red circles (BR 2480). The black line for protons and green line for antiprotons represent the corresponding vLIS's at 122 au. Presently, published BR averaged or 6-months or less period averaged observations of antiprotons are unavailable for comparison. 
\label{fig5} }
\end{figure*}

Figure \ref{fig5}(a) shows the model simulated GCR electron and positron spectra at the Earth for two distinctive periods, May 15-June 11, 2011 (BR 2426) and May 12-Jun 08, 2015 (BR 2480) in comparison with AMS observations of corresponding BRs as indicated. The thick solid lines are the vLIS's for GCR electrons (black line) and positrons (green line), adopted from \citet{2015ApJ...810..141P} and \citet{2019ApJ...873...70A} respectively. The modulated spectra indicate the simulated total modulation at the Earth (1 au) with respect to the corresponding vLIS's at 122 au, where the HP is located. Here the differential intensity, in units of particles m$^{-2}$ sr$^{-1}$ s$^{-1}$ GeV$^{-1}$, is shown as a function of kinetic energy in GeV.

Figure \ref{fig5}(b) shows the model simulated GCR proton and antiproton spectra at the Earth for May 15-June 11, 2011 (BR 2426) and May 12-Jun 08, 2015 (BR 2480) in comparison with AMS observations of corresponding BRs. The black and green lines indicate the corresponding VLIS's for protons and antiprotons at 122 au, adopted from \citet{2015ApJ...815..119V} and \citet{2021Potgieter} respectively. In this panel, the differential intensity in units of particles m$^{-2}$ sr$^{-1}$ s$^{-1}$ GV$^{-1}$ is shown as a function of rigidity. 
Presently, BR averaged observations for antiprotons are unavailable; such observations will be most useful for comparative studies and validation attempts of the antiproton modeling result. In this context,
\citet{2016PhRvL.117i1103A} mentioned the obstacles in measuring antiprotons and that such observations are limited to 1 for about 10$^{4}$ protons. \citet{2013JETPL..96..621A} mentioned that antiprotons are mainly secondary particles produced by interaction of high-energy cosmic hadrons with interstellar matter, but cannot eliminate the possibility of source-like production at shock fronts during supernova explosions; see e.g. \citet{1997ApJ...487..415B}.

It follows from Figure \ref{fig5} that for BR 2426 the modulated electron and positron spectra at the Earth both exhibit a peak just below 1 GeV whereas for protons it occurs between 1-2 GV, in contrast to the modulated anti-proton spectra where it occurs around 3-4 GV. The latter follows evidently because the antiproton vLIS has a vastly different shape in terms of rigidity than the proton vLIS. For the later BR, with lower spectra because of increased modulation, the peak values systematically shift up in rigidity for all the GCRs. Another conspicuous difference in spectra at lower rigidity is that in the case of the modulated electrons and positrons, adiabatic cooling is essentially negligible making diffusion the dominant process below about 100 MeV, whereas for protons and antiprotons the adiabatic cooling is dominant at these low rigidity, causing these spectra to have a similar spectral slope below 1 GV. It is noted that the observed spectra from May 2011 to May 2015 are very well reproduced with an identical set of modulation parameters used for protons and antiprotons, and for electrons and positrons. Only their drift patterns are different, and, of course, their corresponding vLIS's at 122 au.
    
After computing each and every spectrum for protons, electrons, positrons and antiprotons for BRs 2426 to 2480 (May 2011 to May 2015) we set out to fine-tuning these simulated fluxes based on comparison with available observations; for protons with a rigidity bin 1.0-1.16 GV, for electrons and positrons for a kinetic energy bin 1.0-1.22 GeV. This is done for the $A<0$ cycle before the polarity reversal, then for the reversal period without a well define polarity, and for the $A>0$ cycle after the reversal. The purpose of this method was to establish the relative trends and importance of the underlying physics responsible for the modulation of these GCRs, especially over the polarity reversal period, as will be shown and discuss below.

\begin{figure*}
\gridline{\fig{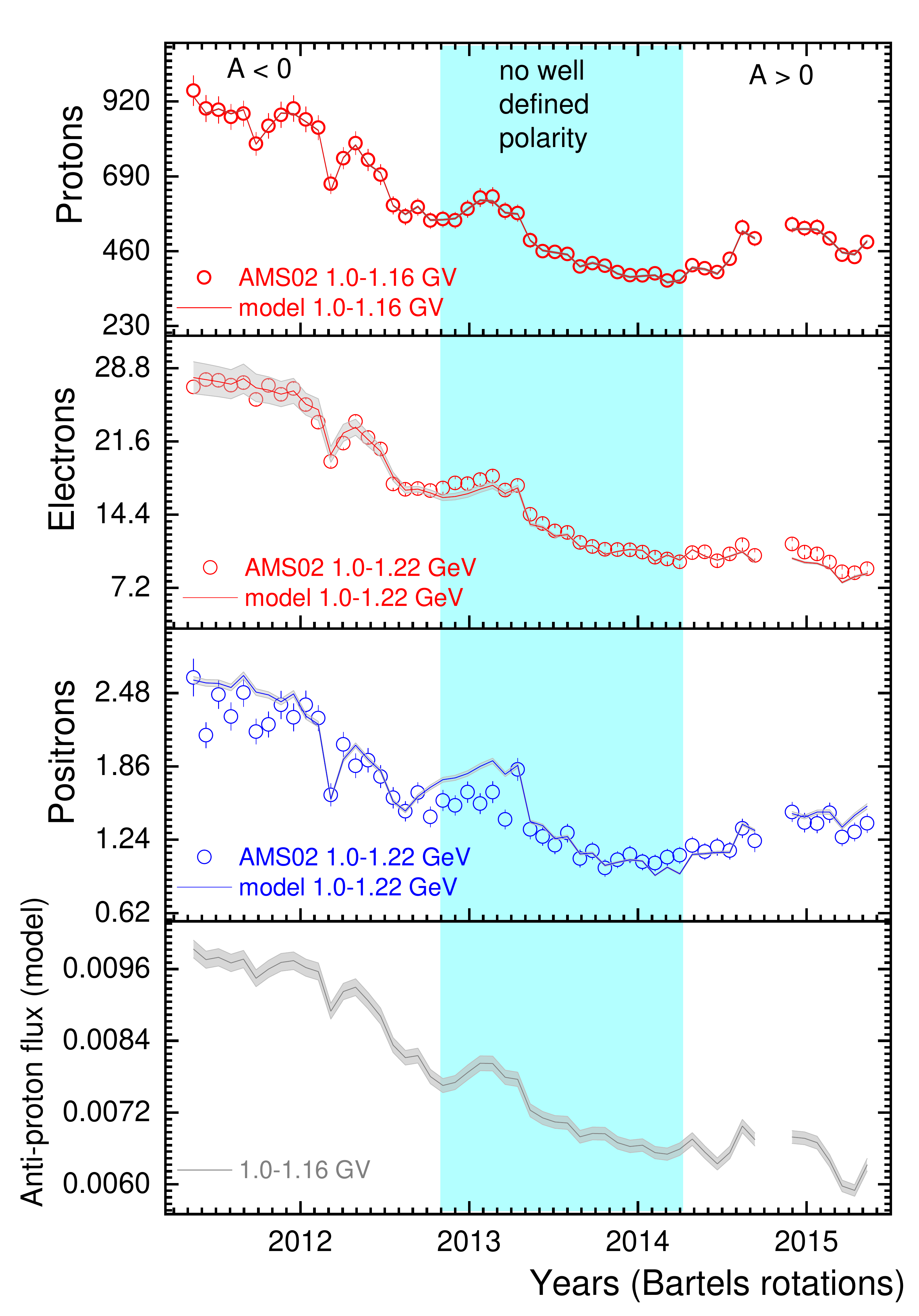}{0.5\textwidth}{(a)}
       \fig{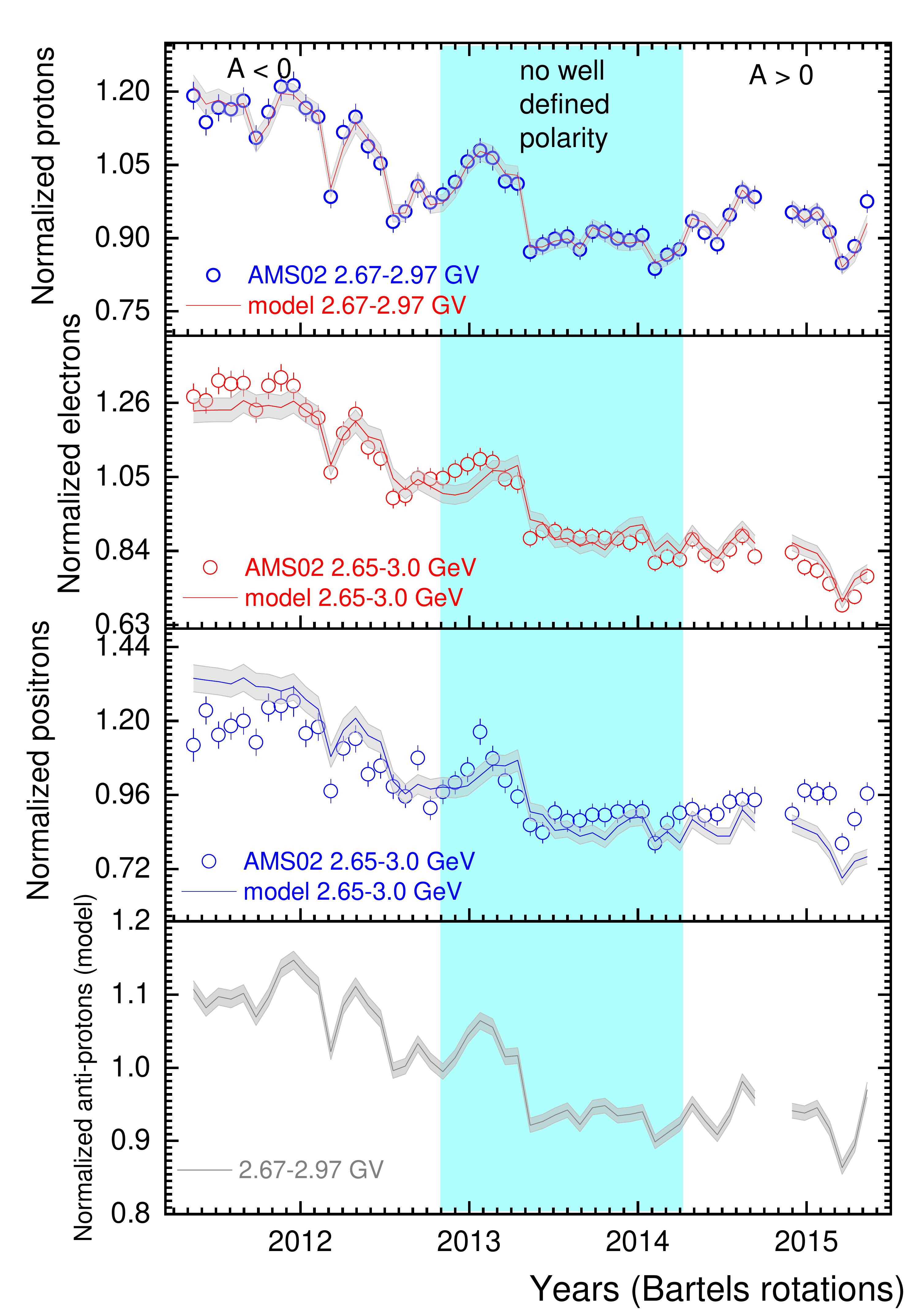}{0.5\textwidth}{(b)}
           }
\caption{Panel (a): Top left panel: Model simulated GCR proton flux for the rigidity range 1.0-1.16 GV (solid line) along with the AMS proton flux (circles) of the same rigidity range. Second left panel: Simulated GCR electron flux over the kinetic energy range 1.0-1.22 GeV (solid line) along with AMS electron flux (circles) of the same kinetic energy range. Third left panel: Simulated positron flux over the same kinetic energy range as electrons (solid line) along with AMS positron flux (circles) of same kinetic energy range. Left bottom panel: Simulated antiproton flux over the same rigidity range as for protons. 
Panel (b): Right panels are similar to lefts panels but now for 2.67-2.97 GV for protons and antiprotons and 2.65-3.0 GeV for electrons and positrons.
The normalization is done based on the average value for the period May 2011 to May 2015. The shaded bands along and around the modeling results (all solid lines) indicate the standard error of mean (standard deviation /$\sqrt n$). 
All plots are shown from May 2011 to May 2015 (BRs 2426 to 2480) with observations missing for BRs 2472 and 2473. Observations are from \citet{2018PhRvL.121e1101A, 2018PhRvL.121e1102A}.   
\label{fig6}}
\end{figure*}

In Figure \ref{fig6}(a) the model simulated GCR proton flux for 1.0-1.16 GV is compared with AMS observations of the same rigidity bin over the period May 2011 to May 2015. Similarly, the model simulated electron and positron fluxes of 1.0-1.22 GeV are compared with AMS observations of same range. In the bottom panel only the simulated antiproton flux for 1.0-1.16 GV is shown. All four particle fluxes considered here show similar variations from May 2011 on wards as the Solar Cycle progresses and all fluxes decrease up to the end of the reversal period. The first noticeable difference between the flux trend of these oppositely charged GCRs becomes evident after the reversal phase when the new A$>$0 epoch starts. The positively charged protons and positrons start to increase slowly, but the negatively charged electrons and anti-protons either decrease (antiprotons) or remain almost the same (electrons). This difference is considered as a clear indication that the drift pattern of these positively and negatively charged GCRs have slowly changed after the polarity reversal of Sun's magnetic field, surely not during the reversal period. 

Figure \ref{fig6}(b) shows the simulated flux variations from May 2011 to May 2015 but now for 2.67-2.97 GV for protons and antiprotons, and 2.65-3.0 GeV for electrons and positrons. The modeling results and AMS observations, for the exact same rigidity and kinetic energy bins, are now normalized with respect to the average of May 2011 to May 2015 (BRs 2426 to 2480) which provides a common reference level for comparison and to better assess the relative changes in the fluxes over time. 
As expected, the particle fluxes at this rigidity level also show similar patterns over time as seen in Figure \ref{fig6}(a).

We note that the proton observations at $\sim$ 3.0 GV show that these fluxes have an overall decrease of $\sim$40\% from May 2011 to February 2014. Electrons and positrons show a decrease of $\sim$40-45\% for this period at $\sim$3.0 GeV but the simulated antiproton flux shows a decrease of only $\sim$20-25\% up to February 2014. 
The recovery after the reversal period seems temporarily interrupted by a transient solar event causing a decrease in April 2015 which is noticeably different for electrons and positrons. Obviously, continued investigate for this whole recovery phase with additional published data are to be done.

\subsection{Modulation over the polarity reversal} \label{sec:polarity-reversal}

We continue in this section with normalized fluxes for both the modeling results and observations for the same period as above in order to compare the variations for electrons on the same scale to positrons, and protons to antiprotons, to better understand similarities and differences over time. As before this is done for the time before, during and after the polarity reversal. 

\begin{figure*}
\gridline{\fig{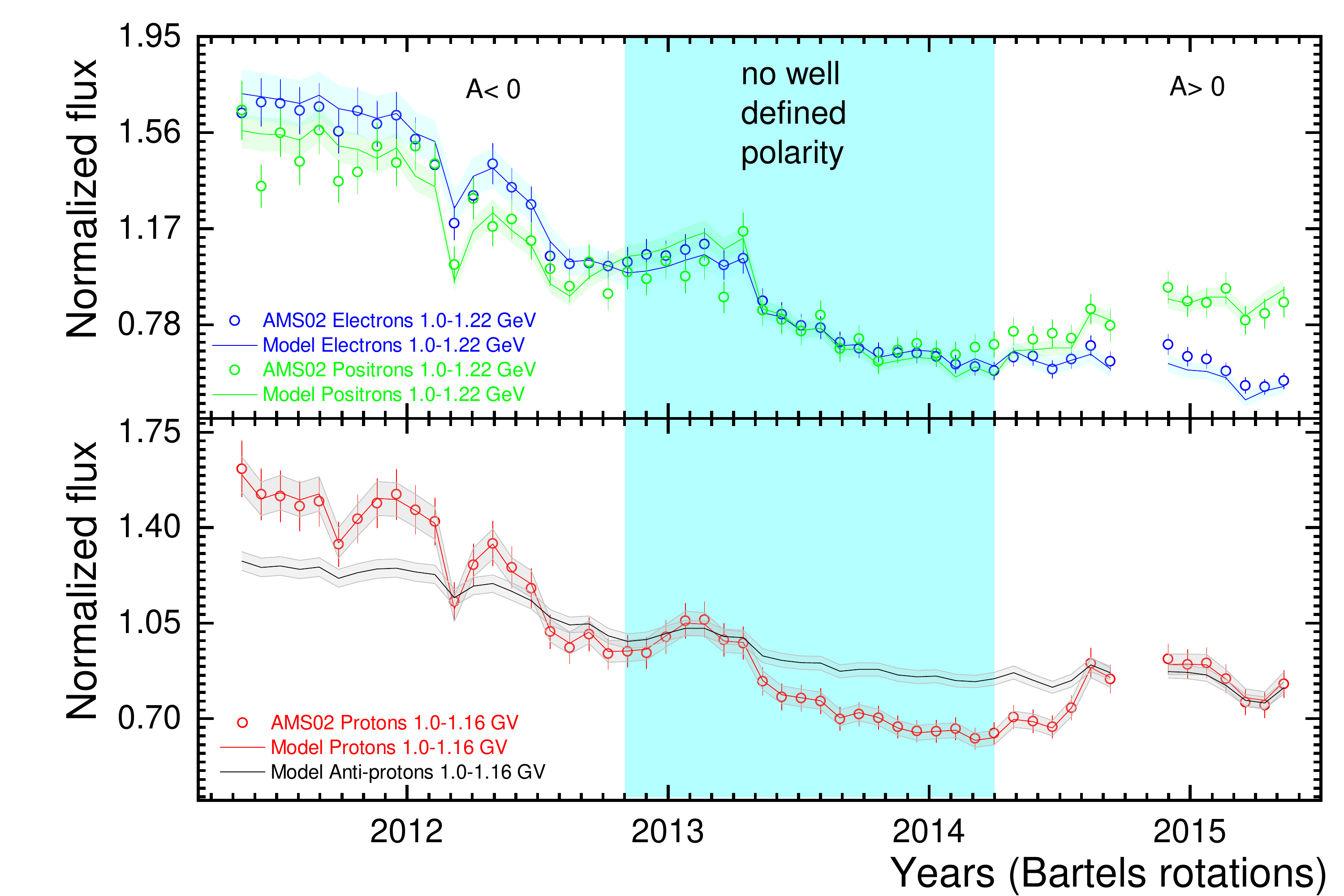}{0.5\textwidth}{(a)}
           \fig{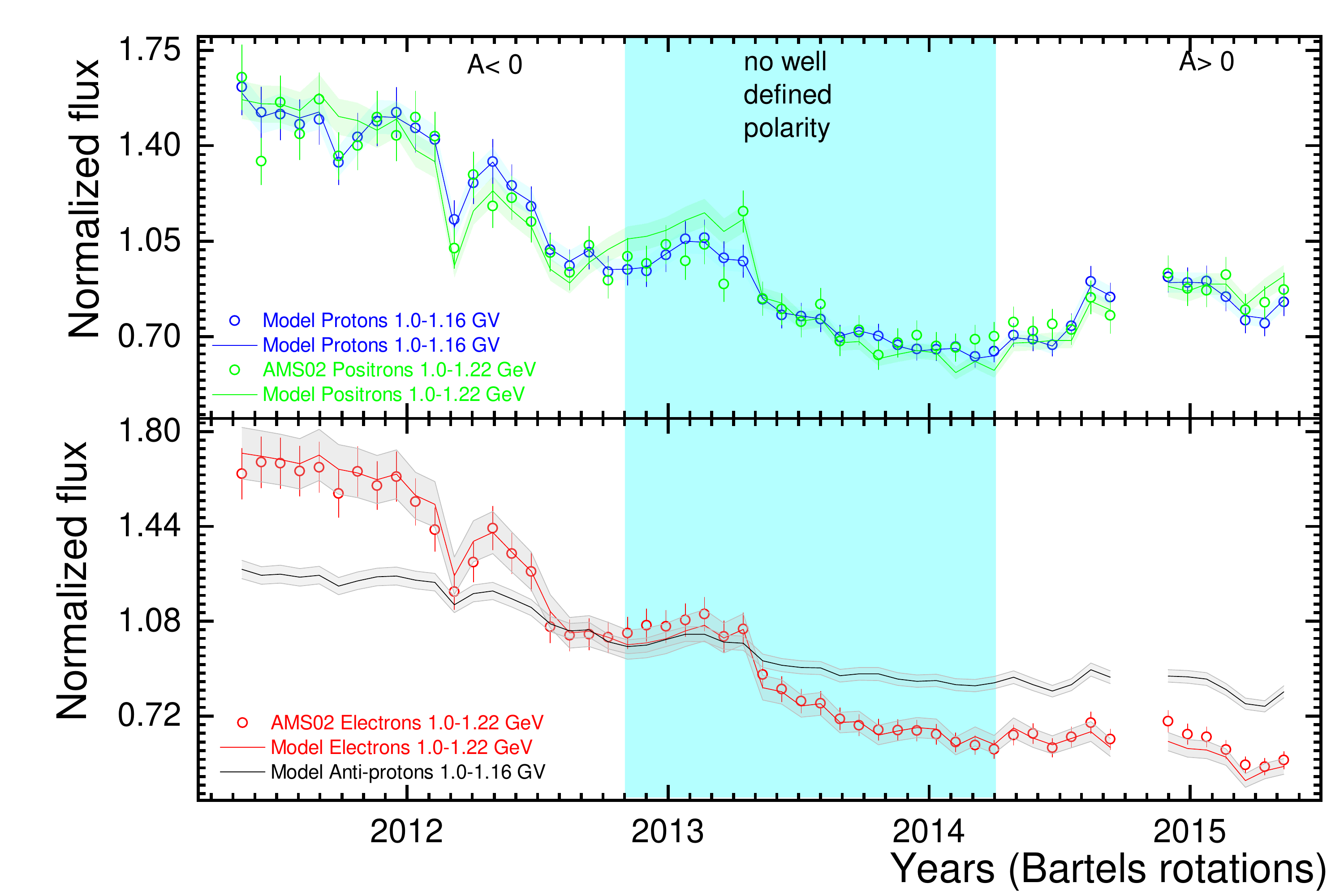}{0.5\textwidth}{(b)}
                   }
\gridline{\fig{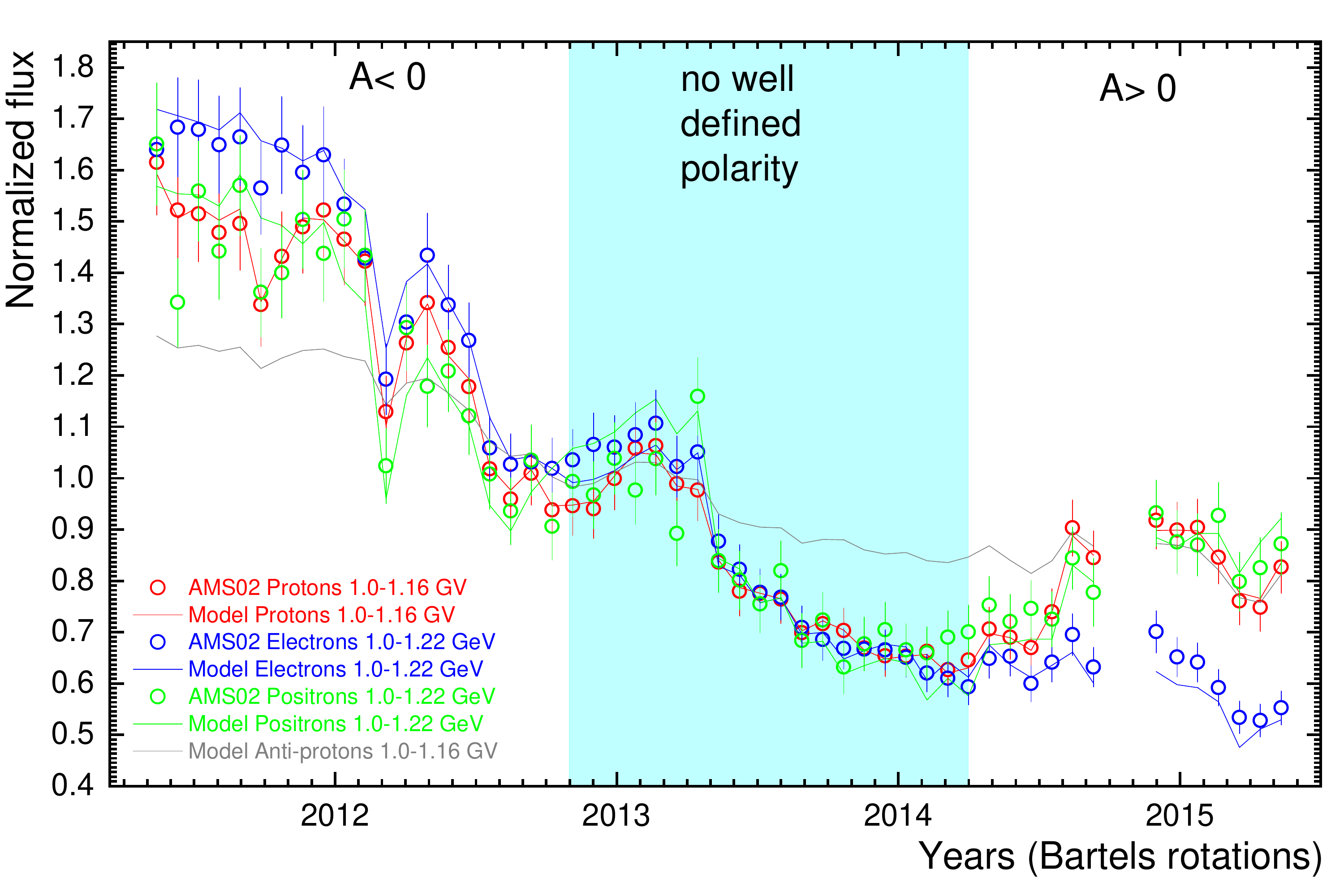}{0.5\textwidth}{(c)}
               }
\caption{Panel (a): Left top panel:  Simulated fluxes (solid lines) normalized for electrons and positrons over 1.0-1.22 GeV in comparison with normalized observed fluxes for electrons (blue symbols) and positrons (green symbols) of the same energy range. Left bottom panel: Normalized, simulated fluxes for protons (in red) and antiprotons (in black) over 1.0-1.16 GV, compared only with normalized proton observations. The period covered is May 2011 to May 2015 (BRs 2426 to 2480) with normalization done with respect to the averaged fluxes from May 2011 to May 2015. As before, the shaded portions indicate the polarity reversal period (November 2012 to March 2014). Error bands to the simulated fluxes are the same as in the previous figure.
Panel (b): Right top panel: As in (a) but now for proton and positron fluxes for 1.0-1.16 GV and  1.0-1.22 GeV, respectively, and compared with AMS proton and positron observations for the same range. 
Right bottom panel: As in the top panel but now for electron and antiproton fluxes for 1.0-1.22 GeV and 1.0-1.16 GV, respectively.
Panel (c): A combination of all the simulated and observed fluxes from (a) and (b) for an easy comparison.   
\label{fig7}}
\end{figure*}

Figure \ref{fig7}(a) shows the normalized AMS electron and positron fluxes at a kinetic energy range of 1.0-1.22 GeV from May 2011 to May 2015 in the top panel, and the normalized proton and antiproton fluxes of the rigidity range 1.0-1.16 GV in the bottom panel. These fluxes and trends are compared to our simulated results for the same range and time. For these simulations all the physical modulation processes are identical for electrons and positrons, except particle drift; the same for protons and antiprotons. It is well-known that during A$<$0 polarity phases, for example, electrons enter the inner heliosphere mainly through the polar regions, while positrons enter mainly along the wavy HCS, thus being affect more by time changes in tilt angles (solar activity). In the top left panel the effect of the change in drift pattern can be noticed from the different trends between electrons and positrons from before to after the reversal period. 
The positron recovery after solar maximum is clearly faster than electrons over the same period.
The lower left panel shows variation of protons and antiprotons. At first, both proton and antiproton fluxes decrease; protons show $\sim$60\% decrease up to October 2012 from May 2011, while the simulated antiproton flux shows only $\sim$25\% decrease during this same period. The only difference between the simulations for protons and antiprotons is their drift direction. Similar overall trends, as noted above, follow from the comparison between proton and antiproton modulation.  
In Figure \ref{fig7}(b) the normalized observed and simulated fluxes for protons and positrons are compared to illustrate that the trends in time are very similar.
For the normalized electron and and antiproton fluxes in the right bottom panel, the overall trends are also similar but the variations are significantly smaller for the antiprotons. Evidently, the differences shown in Figure \ref{fig7}(a) are not present here.
Both protons and positrons start to increase after the completed polarity reversal, but electrons and antiprotons do not begin a recovery until May 2015.
In Figure \ref{fig7}(c) the modeling results and observations shown in panels (a) and (b) are combined in order to make a direct comparison easier.  

\begin{figure*}
\gridline{\fig{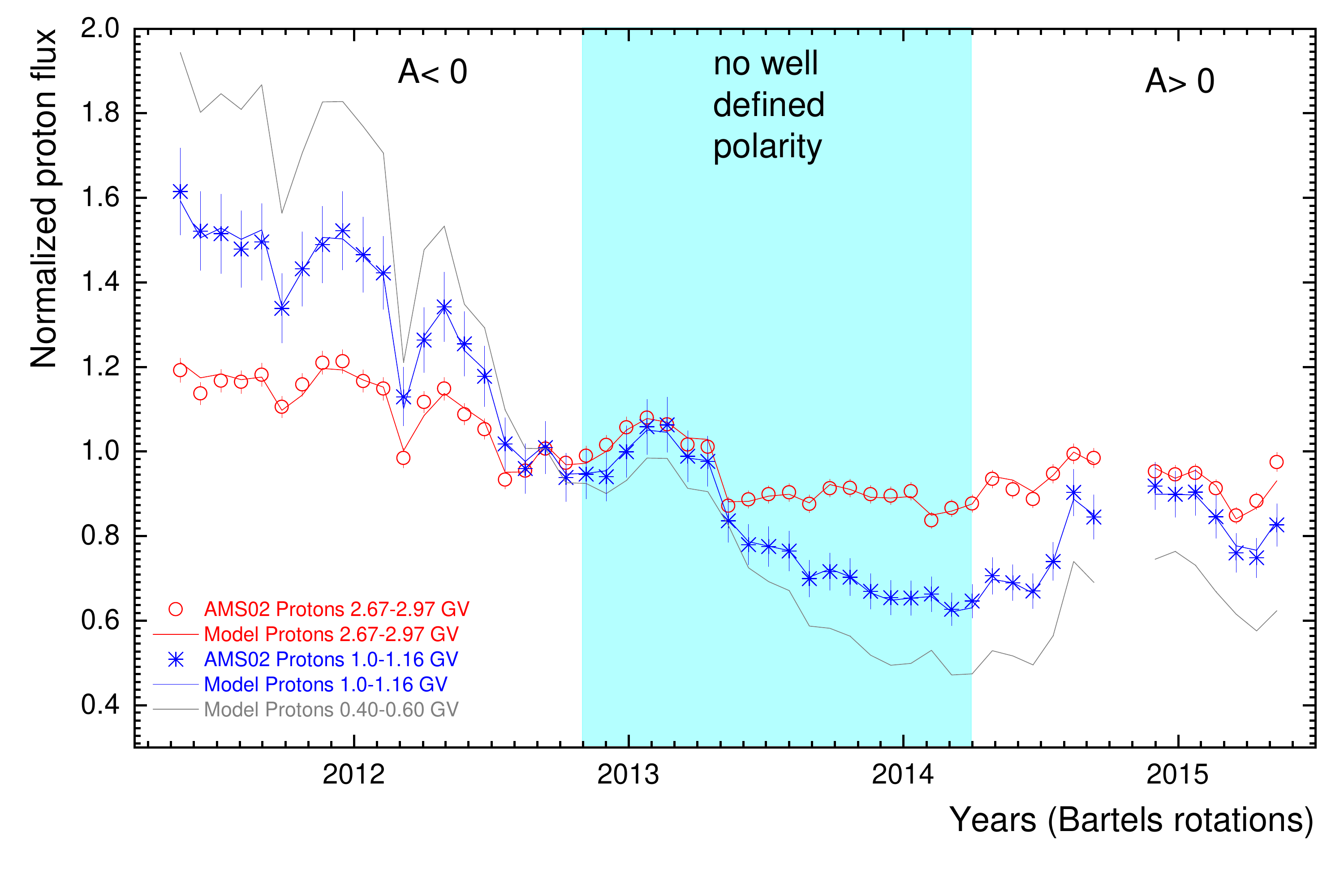}{0.5\textwidth}{(a)}
           \fig{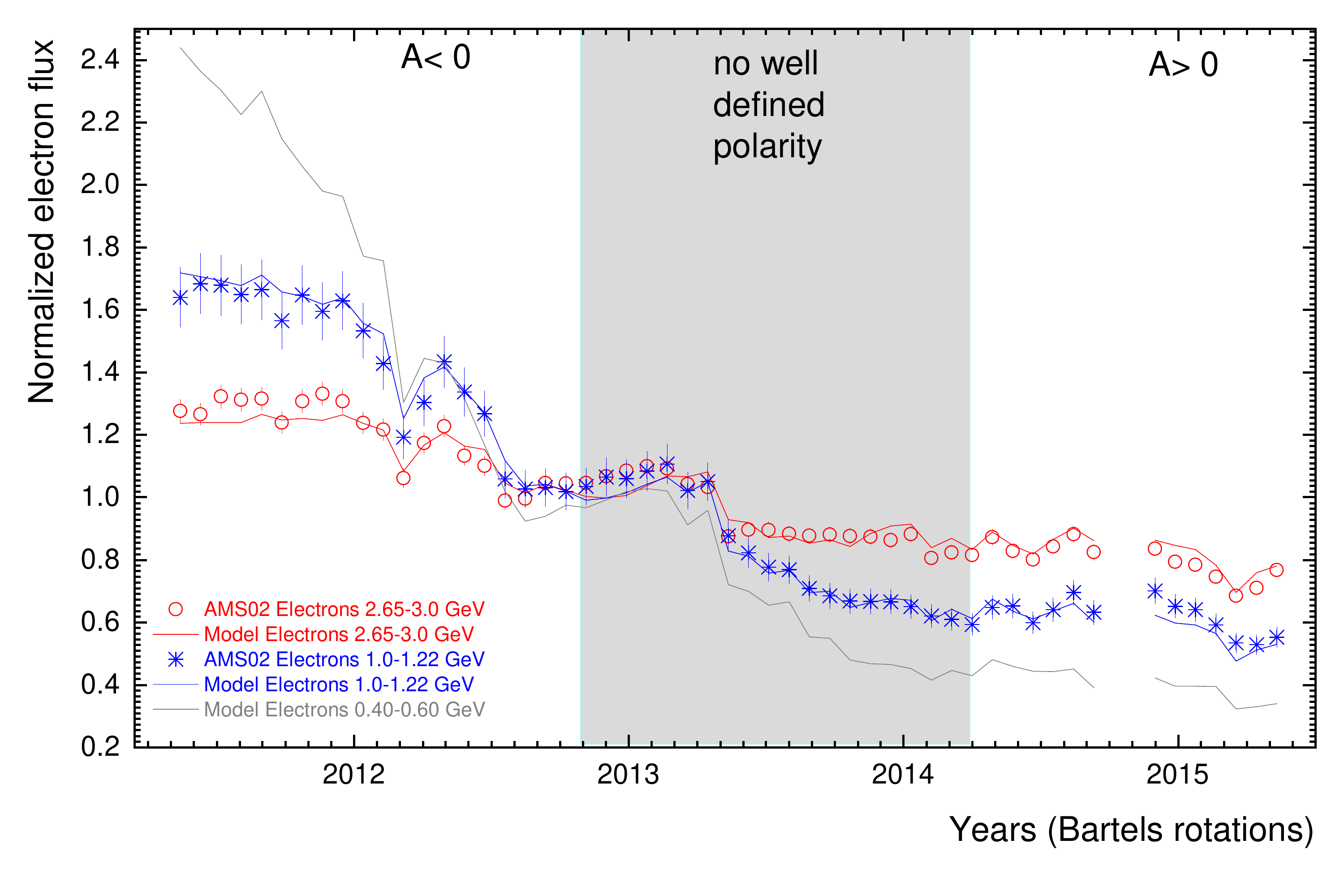}{0.5\textwidth}{(b)}
                    }
\gridline{\fig{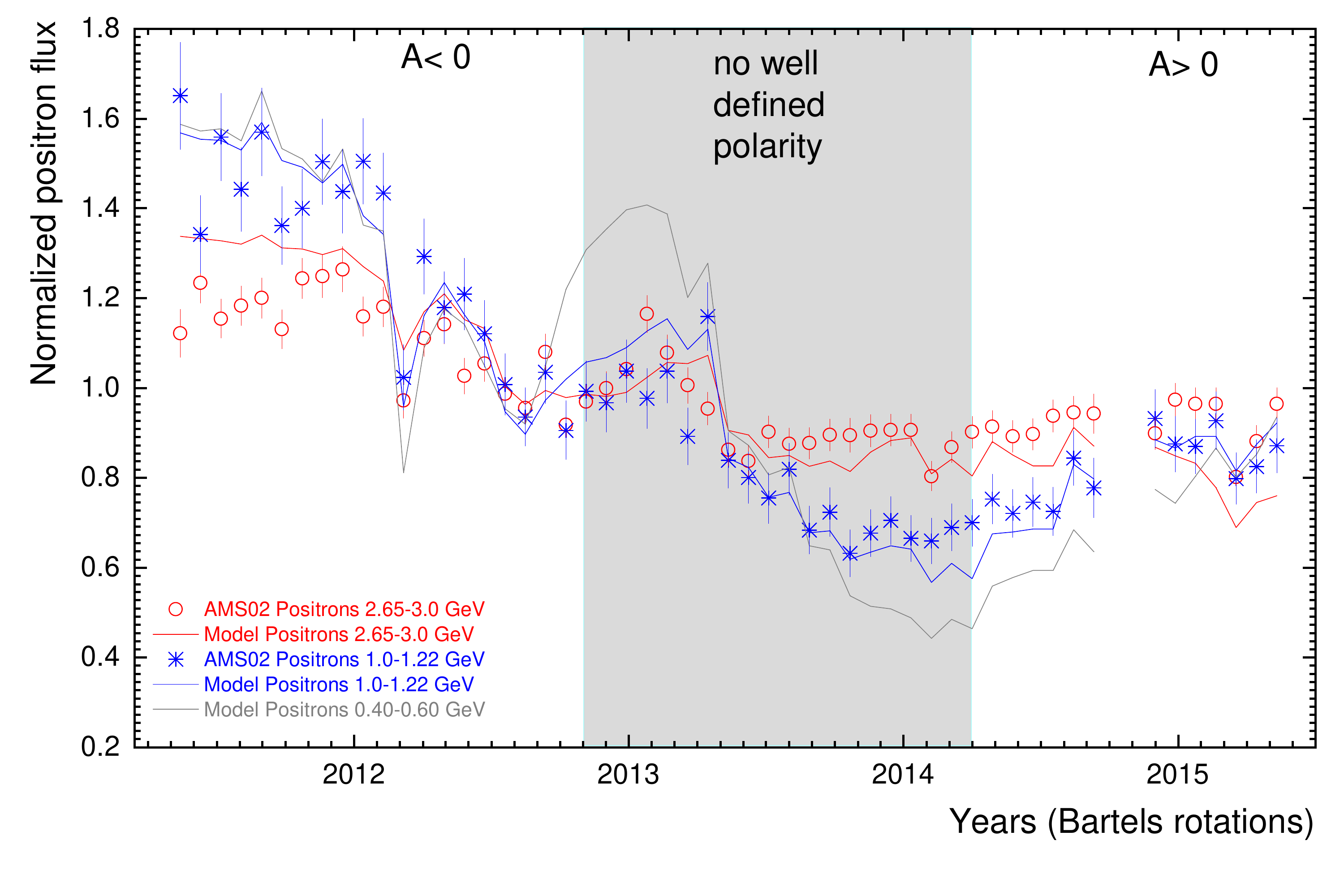}{0.5\textwidth}{(c)}
            \fig{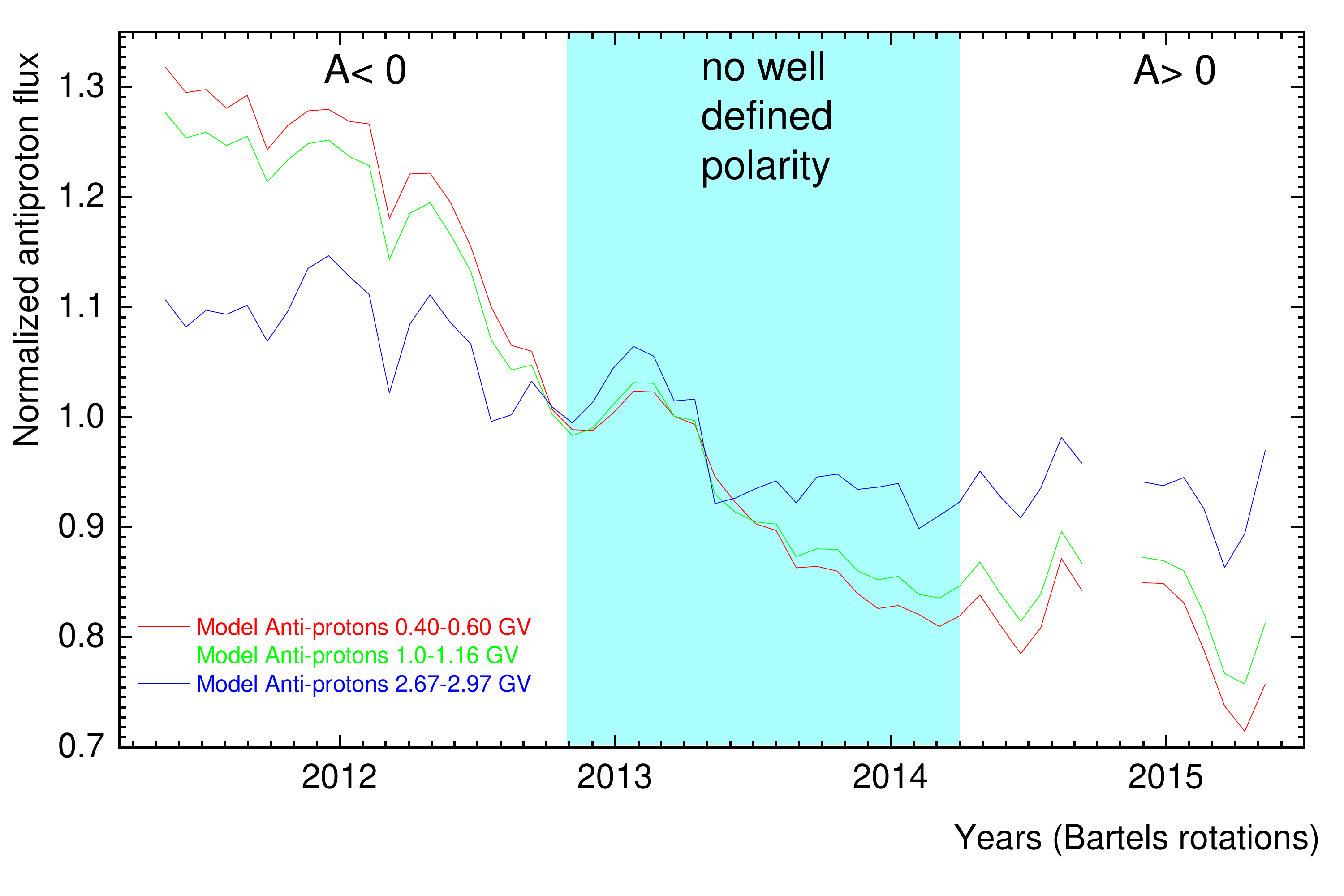}{0.5\textwidth}{(d)}
         }
 \caption{Panel (a): Simulated and normalized proton flux of 0.40-0.60 GV (gray line), 1.0-1.16 GV (blue) and 2.67-2.97 GV (red) are compared with AMS proton observations of rigidity ranges 1.0-1.16 GV (blue) and 2.67-2.97 GV (red). 
 Panel (b): Similar to (a), now for the electron fluxes of 0.40-0.60 GeV (gray line), 1.0-1.22 GeV (blue) and 2.65-3.0 GeV (red), compared with AMS electron observations of 1.0-1.22 GeV (blue) and 2.65-3.0 GeV (red). 
 Panel (c): Similar to (a), now for positron fluxes of 0.40-0.60 GeV (gray line), 1.0-1.22 GeV (blue) and 2.65-3.0 GeV (red), compared with AMS positron observations of 1.0-1.22 GeV (blue) and 2.65-3.0 GeV (red). 
 Panel (d): Similar to (a), now for antiproton fluxes for 0.40-0.60 GV (gray line), 1.0-1.16 GV (blue) and 2.67-2.97 GV (red). All fluxes are shown from May 2011 to May 2015, with normalization done with respect to the May 2011 to May 2015 average.
 \label{fig8}}
\end{figure*}

The next part is focused on comparing simulated fluxes for three different rigidity levels, adding a lower range for which no AMS observations exist. This is done to emphasize the increased flux variations at lower rigidity.
First, for protons shown in Figure \ref{fig8}(a) are normalized, simulated fluxes for 0.40-0.60 GV, for 1.0-1.16 GV and 2.67-2.97 GV, respectively. As before, these computed fluxes are compared to normalized AMS proton observations if available. 
As expected, the total amount of modulation and the flux variations at the lower rigidity is larger than for the higher ranges. This trend is continued during the reversal phase as well with intensities decreasing further to start the recovery consistently at the end of the reversal period. After the reversal period the higher rigidity particles recover somewhat sooner and faster than the low rigidity particles; in this context of a energy dependent recovery, see also simulations by \citet{1992ApJ...390..661L} and \citet{2022AdSpR..69.2330N}. 
The exercise is repeated in Figure \ref{fig8}(b) for electrons and in Figure \ref{fig8}(c) for positrons for the three kinetic energy ranges as indicated. For the computed simulations the exact same set of modulation parameters are used in the model for these GCRs. Clearly, the reproduction of positron variations is not as good as for electrons, with some exaggerated flux at low rigidity at the beginning of the reversal period and an apparent faster recovery after the reversal which seems rather odd when compared to protons. It is unclear to us why this is the case. However, when inspecting the spectra in Figure \ref{fig3}(a,b) it follows that the lower the rigidity is, the larger the modulated difference may become at Earth between these GCRs. It is expected that the electron to positron flux ratio will be much different with time at lower rigidity than at the reported rigidity, as discussed by \citet{2021ApJ...909..215A}, and explored below. 
The exercise is completed as shown in Figure \ref{fig8}(d) for antiproton fluxes for 0.40-0.60 GV, 1.0-1.16 GV and 2.67-2.97 GV. 

\begin{figure*}
\gridline{\fig{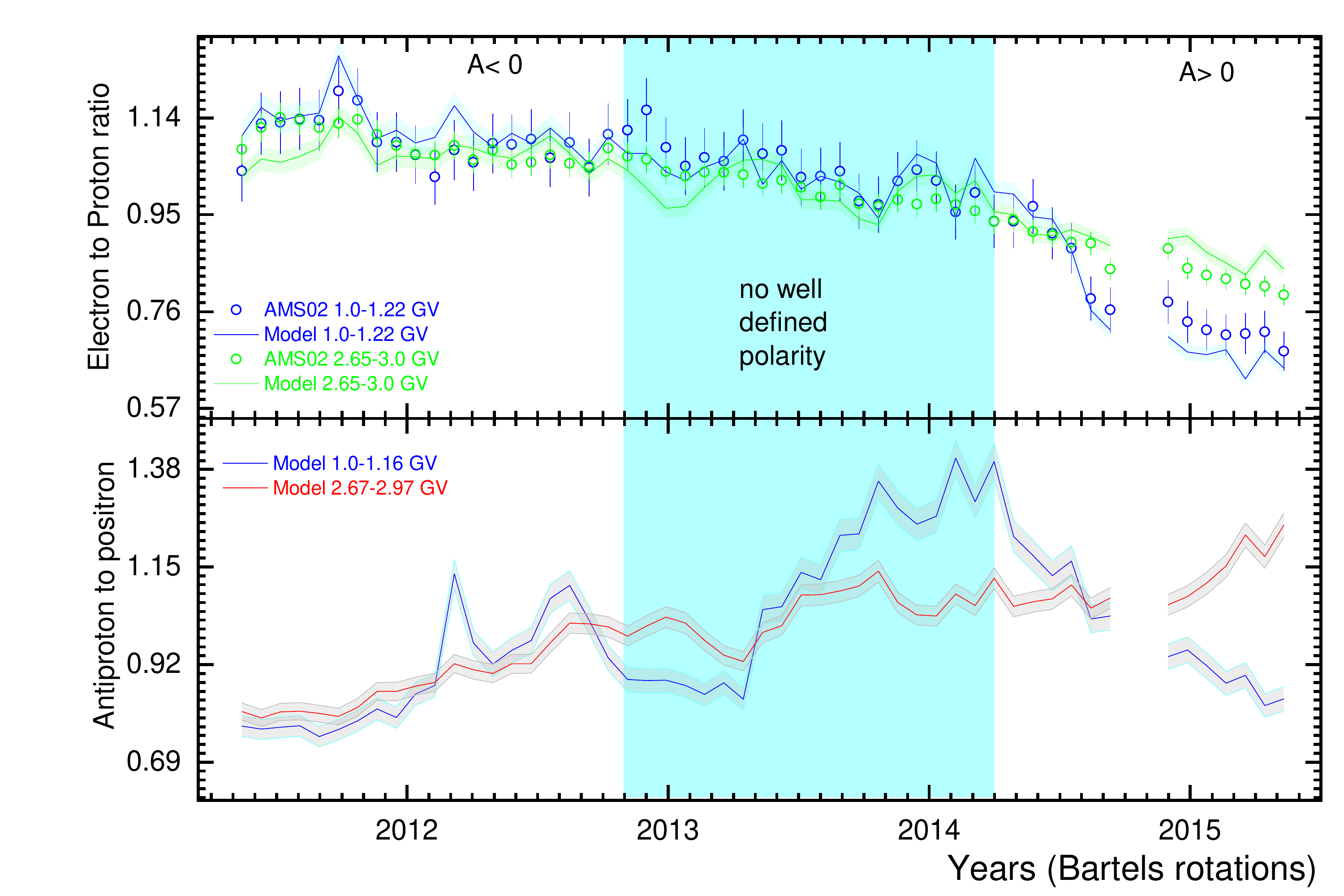}{0.5\textwidth}{(a)}
       \fig{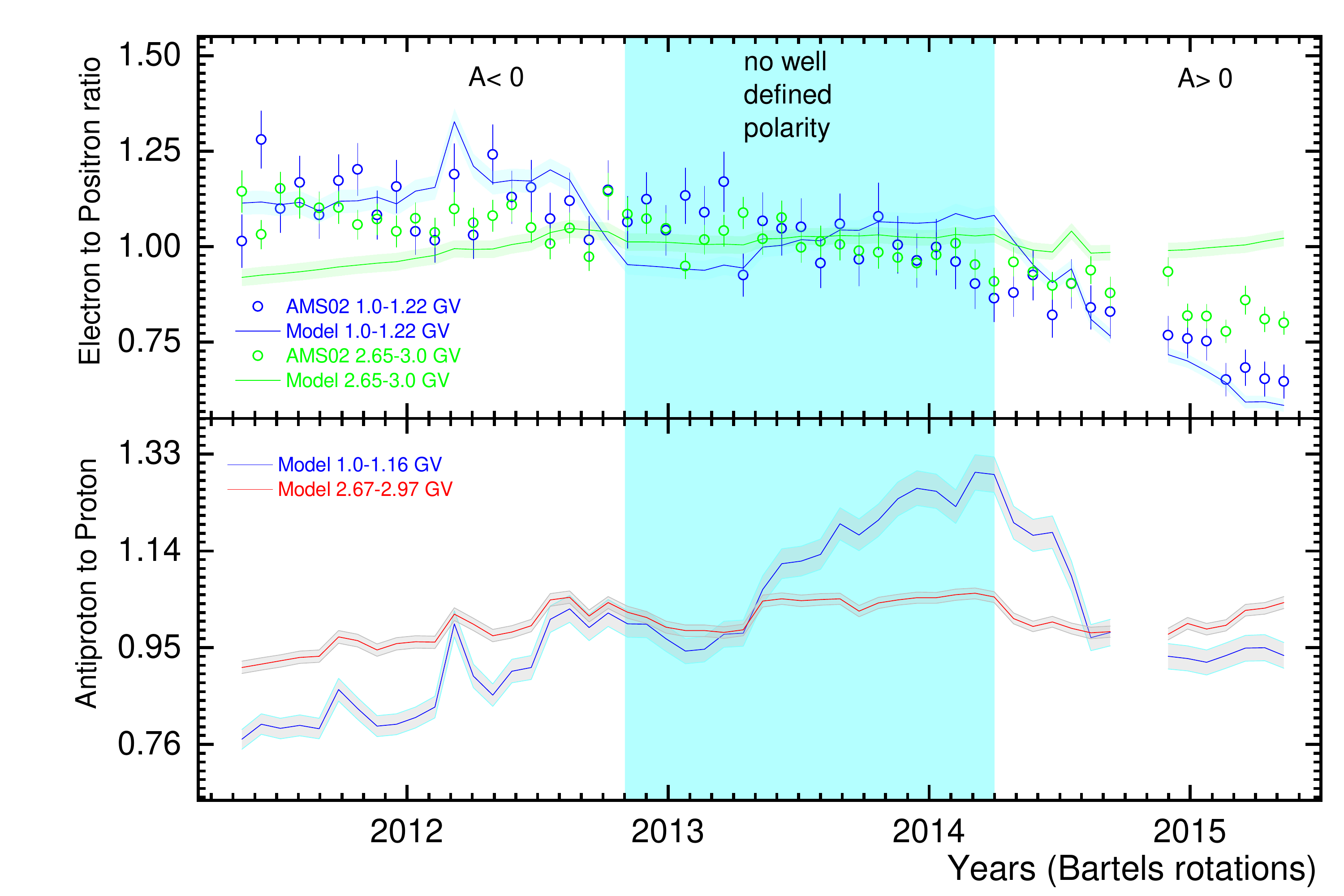}{0.5\textwidth}{(b)}
                  }
    \gridline{\fig{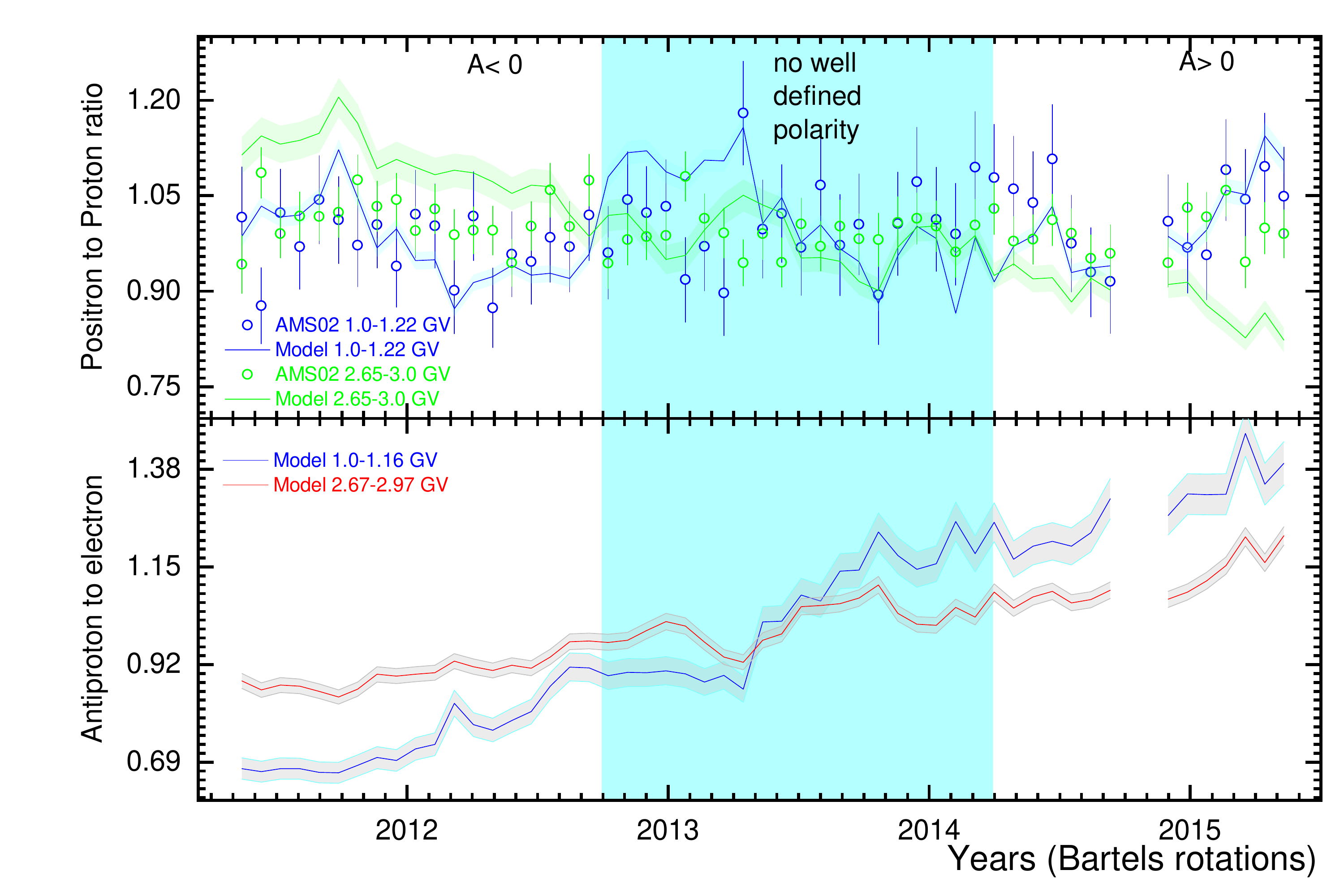}{0.5\textwidth}{(c)}
                }
 \caption{Panel (a): Top left: Computed simulations of the electron to proton flux ratio over rigidity ranges of 1.0 - 1.22 GV (blue) and 2.65 - 3.0 GV (green) compared with corresponding AMS observations (blue and green circles). Bottom left: Computed simulations of the antiproton to positron flux ratio over 1.0 - 1.22 GV (blue) and 2.65 - 3.0 GV (red). (For electrons and positrons kinetic energy $\approxeq$ rigidity).
 Panel (b): Top right: Simulated electron to positron flux ratio for 1.0 - 1.22 GV (blue) and 2.65 - 3.0 GV  (green) compared with AMS02 observations (blue and green circles) of same rigidity bins. Right bottom: Simulated antiproton to proton flux ratio for 1.0 - 1.16 GV (blue) and 2.67 - 2.97 GV (red). 
 Panel (c): Top: Simulated positron to proton flux ratio for 1.0 - 1.22 GV (blue) and 2.65 - 3.0 GV  (green) compared with corresponding AMS02 observations (blue and green circles). Bottom: Simulated antiproton to electron flux ratio for 1.0 - 1.16 GV (blue) and 2.67 - 2.97 GV (green).
\label{fig9}}
\end{figure*}

\subsection{Cosmic ray flux ratios} \label{sec:ratio}

In the following, the flux ratios are shown particularly at two different rigidity/energy ranges and as a function of time, May 2011 to May 2015. The purpose is to highlight what happens to the various flux ratios before ($A<0$ cycle) to after ($A<0$ cycle) the reversal period at the selected rigidity and how drift patterns affected the ratios of oppositely charged particles. 

In Figure \ref{fig9}(a) the electron to proton ratio and the antiproton to positron ratio are shown, respectively in the upper and lower panels. In Figure \ref{fig9}(b) the electron to positron ratio and the antiproton to proton ratios are shown, respectively in the upper and lower panels and in Figure \ref{fig9}(c) the positron to proton and antiproton to electron ratios are shown to complete the picture. Comparisons are made to observations if available.
It follows from Figure \ref{fig9}(a) that the electron to proton ratio at $\sim{1}$ GV decreased 
by $\sim$50\% over the mentioned period but the largest contribution comes from the period after the reversal. At $\sim{3}$ GV the trend is similar but the overall decrease it less. The simulated antiproton to positron ratio at $\sim{1}$ GV exhibits an increase of $\sim$ 50-60\% from May 2011 to the end of the reversal period, followed by a decrease of $\sim$ 50-60\%, but it happened much faster (with in a year) whereas at $\sim{3}$ GV the ratio progressively increases over this period.
We note that before the polarity reversal both electrons and antiprotons drift inward mainly through the heliospheric polar regions while protons and positrons enter mainly along the HCS. There is no significant difference in flux ratio until the polarity reversal starts. As solar activity level increases all the GCR fluxes decrease with the same relative rate; see Figure \ref{fig7} (c); during the polarity reversal period the electron to proton ratio shows a slight decrease as indication of an uncertain (not steady) drift pattern (in preparing for the drift reversal), and a large flux ratio decrease after the completed HMF reversal, now as expected and an indication of the reversal of the drift pattern settling in.  A similar trend is expected for antiprotons to positrons, but the flux ratio shows an increase during the reversal period and a peak is reached around February-March, 2014. After the reversal, the trend in the flux ratio seems the same. It is worth mentioning that apart from particle drift, adiabatic energy change is the physical process that also modulate protons/antiprotons differently from electrons/positrons, especially at lower rigidity where the latter is mostly diffusion dominated.
The peculiar shape of the antiproton vLIS at lower rigidity, especially below $\sim$ 3.0 GV should also be kept in mind. 
In Figure \ref{fig9}(b) the electron to positron ratio at $\sim{1}$ GV decreased $\sim$ 40\% from May 2011 to May 2015, again with the largest contribution to this decrease from after the reversal. At $\sim{3}$ GV the observational trend is similar but the decrease it less, whereas the model predicts a smaller variation over this period. The simulated antiproton to proton ratio at $\sim{1}$ GV follows a similar trend than the antiproton to positron ratio with an increase of $\sim$ 50-60\% from May 2011 to the end of the reversal period, followed by a clear decrease of $\sim$ 40\%, whereas at $\sim{3}$ GV the ratio increases over this period.
Comparing electron to positron flux ratios with that for antiprotons to protons, we again note that a reversal in drift direction must have occurred after the polarity reversal, indicated by the decrease in the ratio in case of electrons to positrons at $\sim$ 1 GV, but the antiproton to proton ratio shows an increase during the reversal phase and decreases only after the completion of the reversal, i.e. the proton flux reduced relatively more at this rigidity.
In Figure \ref{fig9}(c) the positron to proton ratio at $\sim{1}$ GV does not show a consistent long-term trend while varying over shorter time spans. At $\sim{3}$ GV the observational trend again shows no consistent long-term trend but the model predicts a decreasing trend over the full period. The simulated antiproton to electron ratio exhibits an increasing trend at both rigidity values over the full period. 
Since both positrons and protons enter the inner heliosphere mainly along the HCS before the reversal but mainly through the polar regions after the reversal, any noticeable difference is not expected from before to after the reversal but there are some flux variations not being completely identical, which needs further investigation as to the cause of this.
The antiproton to electron flux ratio shows a gradual increasing trend for the whole selected period, which seems strange, but of course, the electron flux at $\sim$ 1 GV does not start the recovery until May 2015, also see Figures \ref{fig6}, \ref{fig8} (b) \& \ref{fig8} (d).

Unfortunately, the AMS observations are reported only for 1 GV and above, whereas large and quite spectacular modulation and drift effects occur at lower rigidity, see e.g. \citet{2003GeoRL..30.8032H,2004JGRA..109.1103L,2004ApJ...602..993P,2016AdSpR..58..453N,2017A&A...601A..23P,2017ApJ...834...89D}.

\begin{figure*}
\gridline{\fig{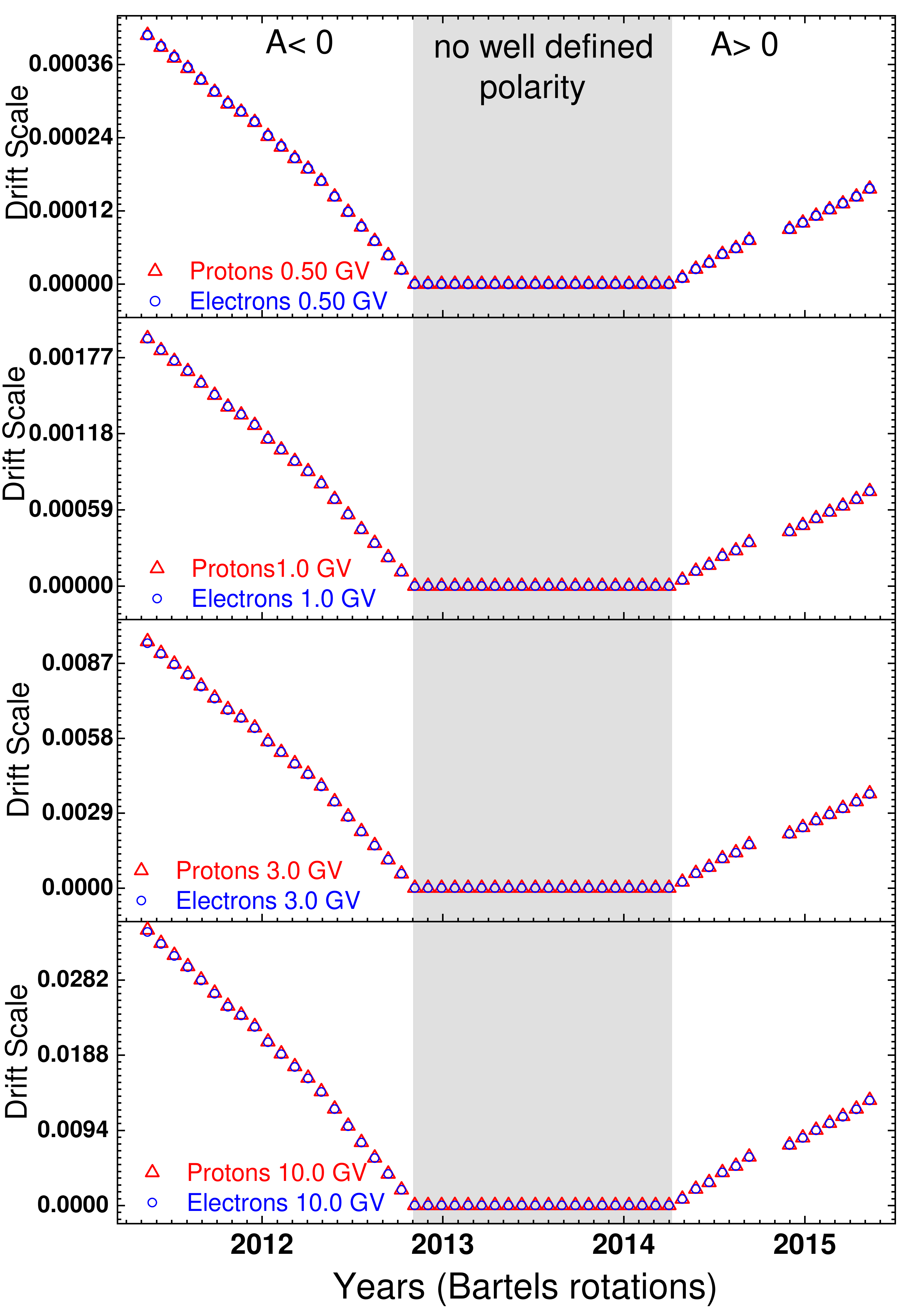}{0.33\textwidth}{(a)}
          \fig{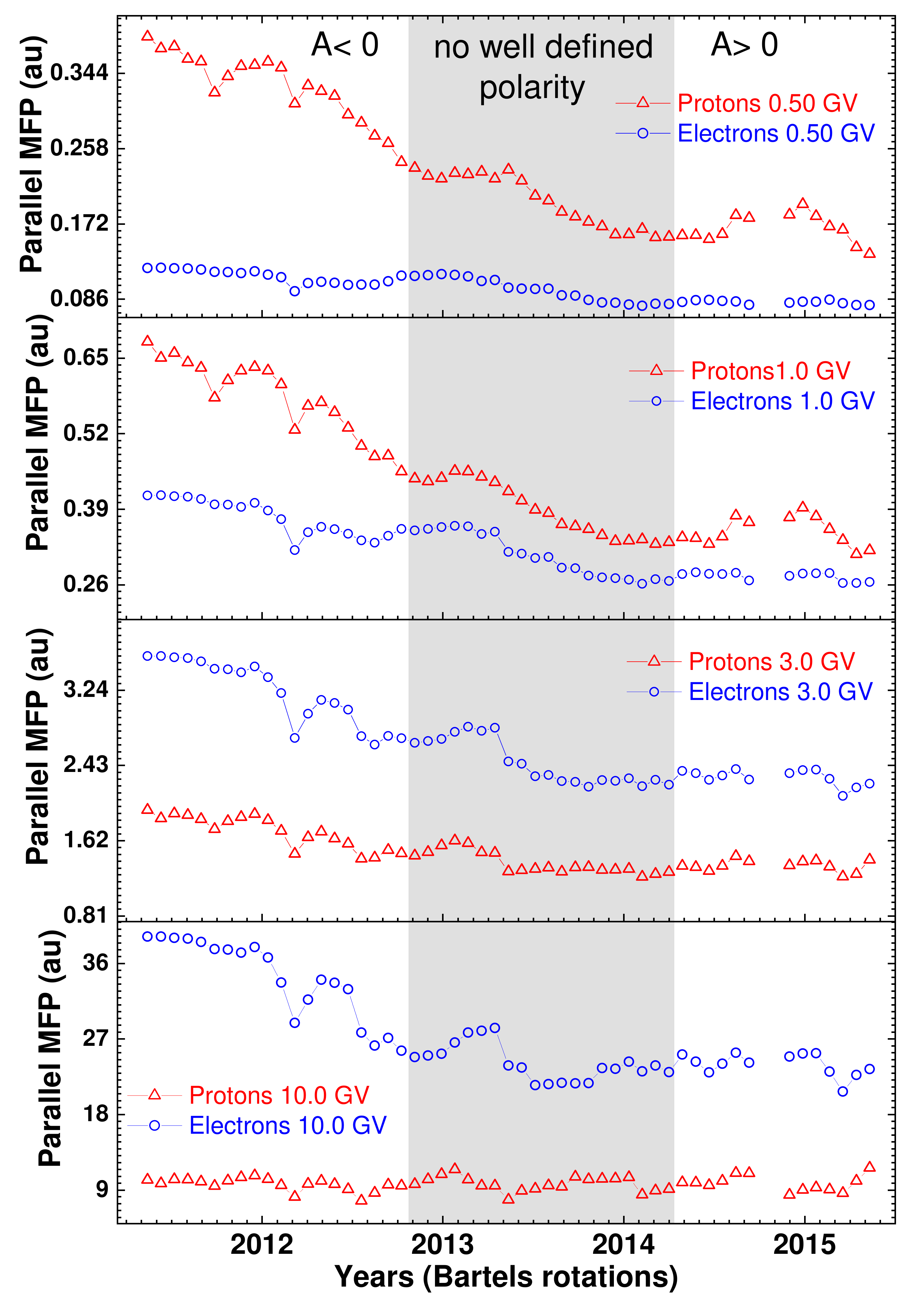}{0.33\textwidth}{(b)}
          \fig{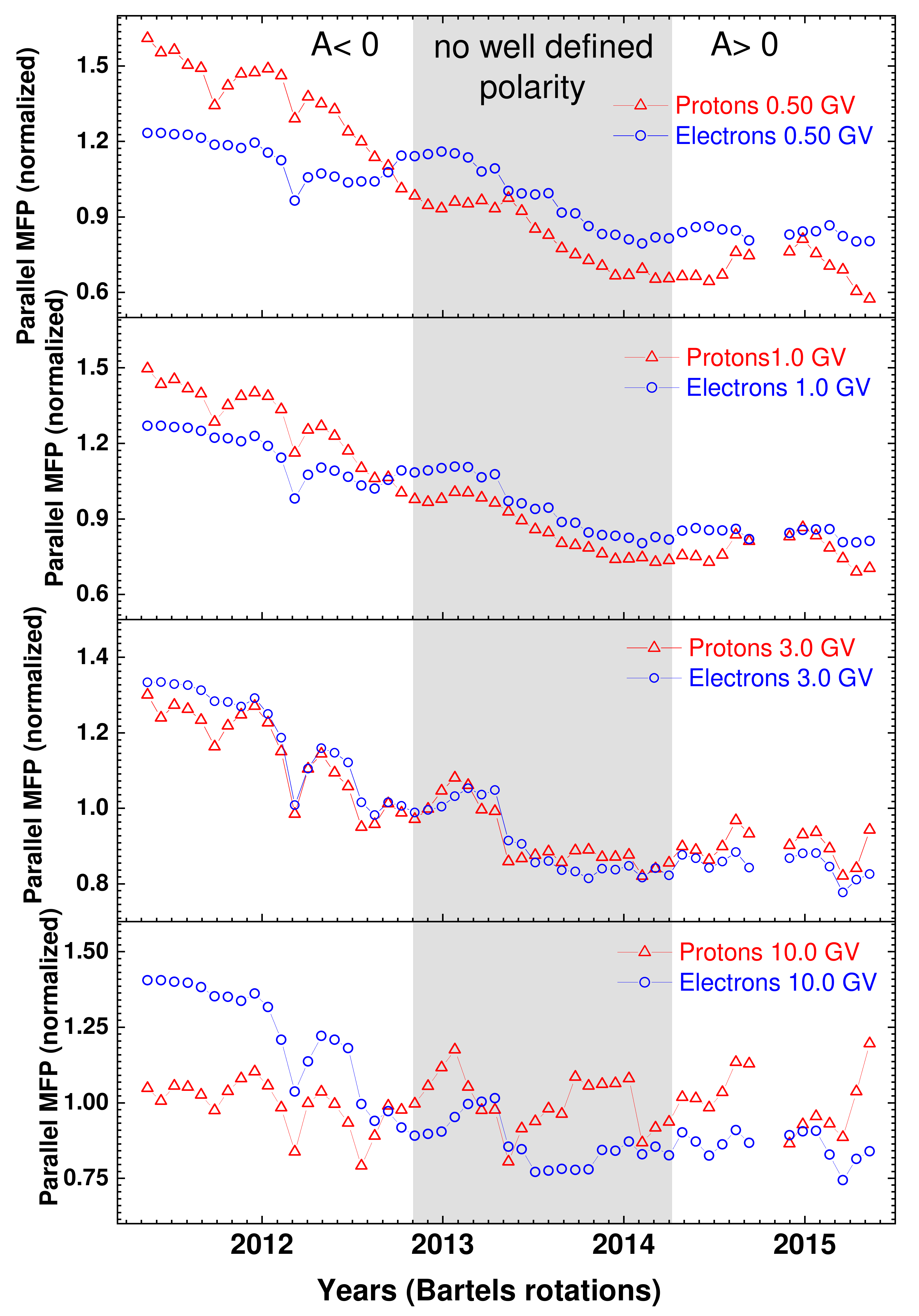}{0.33\textwidth}{(c)}
          }
\caption{Panel (a): Variation of the drift scale (based on Equation \ref{Eq4}) over time, from May 2011 to May 2015, at 0.5 GV (upper panel), 1.0 GV (second panel), 3.0 GV (third panel), and 10.0 GV (bottom panel) for protons (red triangles) and electrons (blue circles). 
Panel (b): Variation of the parallel MFP (see Equation \ref{Eq5}) in units of au over time, from May 2011 to May 2015, for protons and electrons at the same four rigidity values. 
Panel (c): Normalized parallel MFP variation over the same time for protons and electrons at the same four rigidity values. 
The polarity reversal phase is gray shaded.
\label{fig10}}
\end{figure*}

\subsection{Diffusion and drift coefficients over time (solar activity)} \label{sec:DCs}

In this section, the focus is on the physics as represented by the drift and diffusion coefficients used in the model in order to simulate the available observations, in terms of both their rigidity/energy range and time dependence. The purpose is to illustrate the relative variations of the underlying physics for GCR modulation for the period under investigation.

In Figure \ref{fig10}(a) the trends from May 2011 to May 2015, including the reversal period, for the drift scale are shown at four rigidity values, 0.5 GV (top panel), 1.0 GV (second panel), 3.0 GV (third panel), and 10.0 GV (bottom panel) for both protons (red triangles) and electrons (blue circles).  
The drift scale is obviously identical for protons and electrons. However, the emphasis here is on the long-term trends, particularly what happens before, during and after the reversal period. It follows that the changing rate before the reversal is different from after the reversal (less steep) whereas during the reversal it is consistent with essentially zero drift.
The drift scale value is $\sim$5 times higher at 1.0 GV compared to 0.5 GV, similarly $\sim$5 times higher at 3.0 GV compared to 1.0 GV, and $\sim$18 times higher at 10.0 GV compared to drift scale at 1.0 GV. 
Relating what is shown here to Equation \ref{Eq4}, the value of $K_{A0}$ was 0.75 for BR2426 (May-June 2011), then gradually reduces to zero by BR2446 (October-November 2012), stay at zero until the polarity reversal is completed with BR2465 (March-April 2014), and then gradually increase to 0.38 by BR2480 (April-May 2015). The value of $P_{A0}$ in Equation \ref{Eq4} is 0.90 GV for both protons and electrons for this considered period.
Our results show a slight difference in the slope of the decreasing drift scale in 2012, after BR 2437, which could be related to the dips in the GCR flux during this period which could have changed the modulation conditions further out in the heliosphere after this period.  A slight shift in the slope of the increasing drift scale after the polarity reversal is noted in 2014, after BRs 2472-2473; unfortunately for BRs 2472/2473 no AMS02 observations are available. 
In Figure \ref{fig10}(b) the time variation of the parallel MFPs for protons (red triangles) and electrons (blue circles) are shown for the same period and the four rigidity values. In Figure \ref{fig10}(c) this is repeated for normalized MFPs, with respect to the May 2011 to May 2015 average. 
It follows from Figure \ref{fig10}(b) that the values of the MFPs for electrons and protons are different, also in time (with solar activity), but these differences generally become less with increasing rigidity. 
At the lower rigidity values, 0.5 GV and 1.0 GV, we find the value of the parallel MFP for protons to be higher than for electrons; $\sim$3 times higher at 0.5 GV and $\sim$1.5 times higher at 1.0 GV during 2011-2012, and that this difference gradually decreases as solar activity progresses towards maximum, to remain steady after the polarity reversal, until May 2015. At higher rigidity values, 3.0 GV and 10.0 GV, the value of parallel MFP for electrons are higher compared to protons; $\sim$ 2 times higher at 3.0 GV, $\sim$ 2-4 times higher at 10.0 GV. 
At all rigidity values the difference between the parallel MFPs of electrons and protons reduce towards solar maximum activity, during the reversal phase, while after the reversal that difference maintains. It seems the shifting (to higher value for electrons) of the parallel MFP value occurs at a rigidity in-between 1.0-3.0 GV.
Relating to Equation \ref{Eq5}, the value of $(K_{\parallel})_{0}$ is 49.0 (here in units of 10$^{22}$cm$^{2}$s$^{-1}$ and in the rest of the sentence) for May 2011, reducing to 26.49 by the end of polarity reversal phase, and gradually increasing up to 30.52 by May 2015 for protons (and antiprotons). In the case of electrons (and positrons), $(K_{\parallel})_{0}$ is 29.84 in May 2011, reducing to 21.09 by March-April 2014, then slowly increasing back, up to 25.21 until May 2015. 

In Figure \ref{fig10}(c), the normalized parallel MFPs of electrons and protons at different rigidity shows the relative trends. The MFPs decrease until the polarity reversal is completed, but their slope of change is different for protons and electrons before, during and after HMF polarity reversal. This trend is different at lower rigidity (0.5 GV) compared to higher rigidity (10.0 GV). At 3.0 GV the parallel MFPs of both proton and electron have a matching trend for the whole considered period. 
These normalized MFPs show that the short-term trends in time are not always synchronized probably because of different drift directions. 

\begin{figure*}
\gridline{\fig{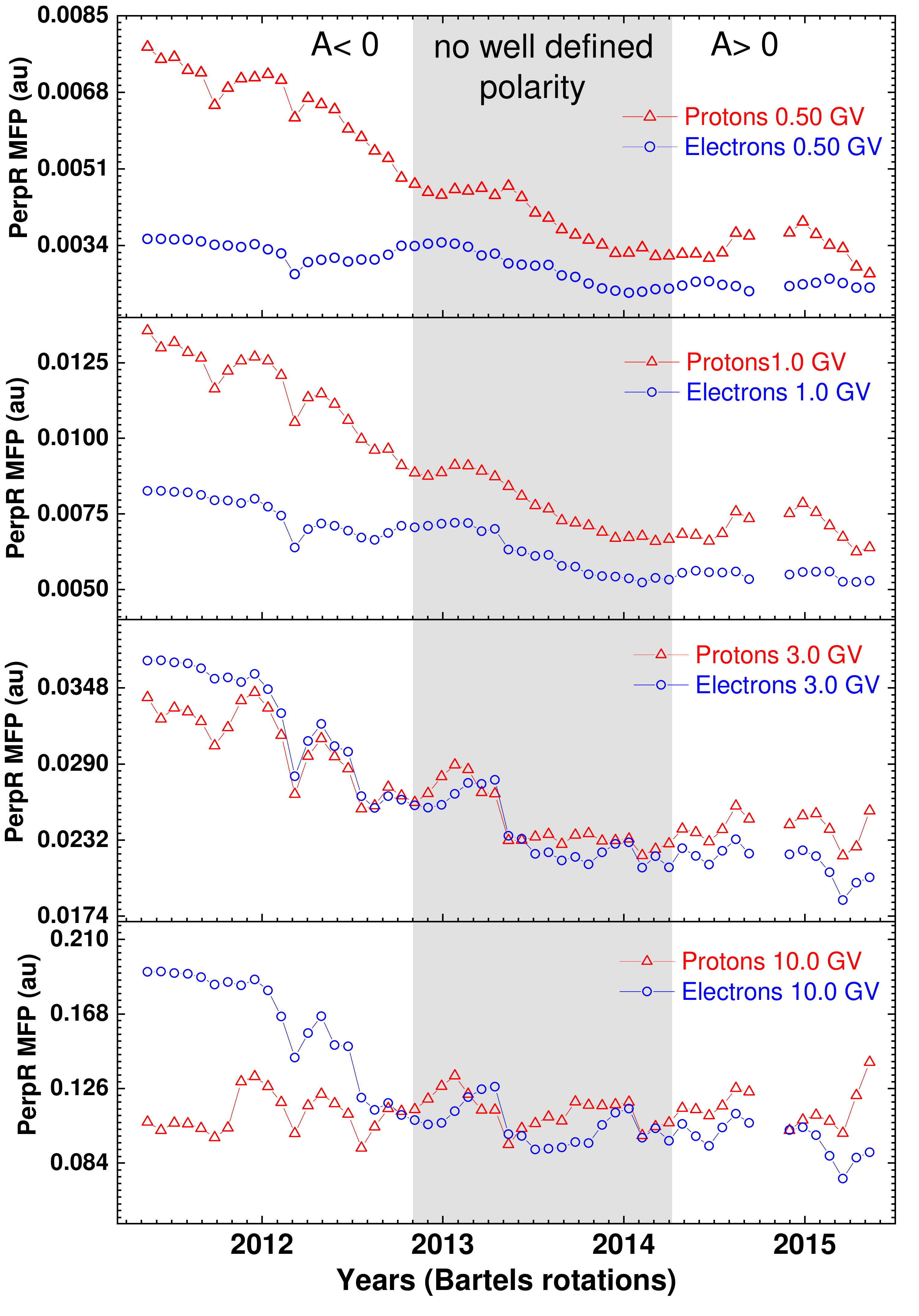}{0.5\textwidth}{(a)}
          \fig{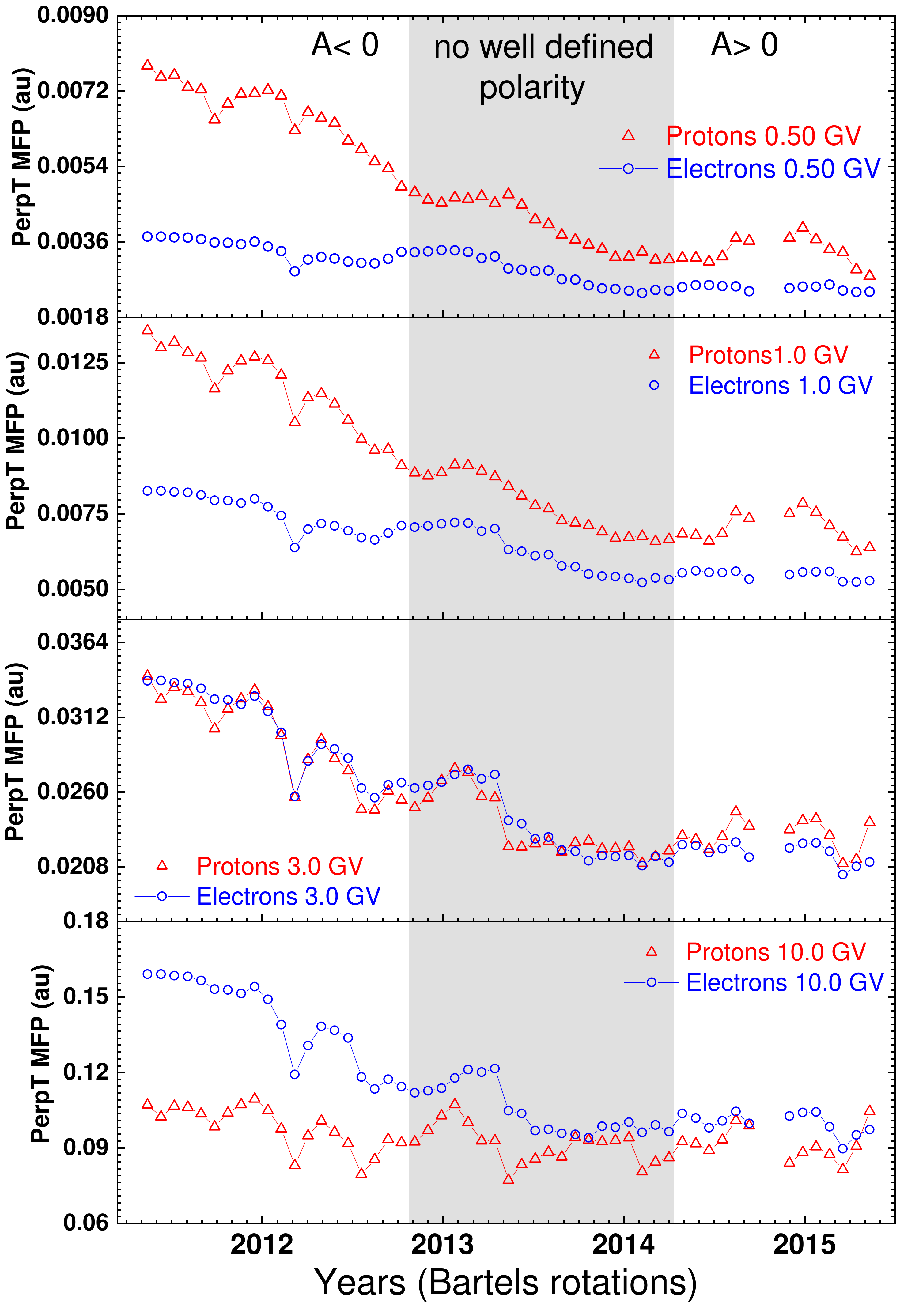}{0.5\textwidth}{(b)}
                }
\caption{Panel (a): Trend variation in the perpendicular MFP in the radial direction (see Equation \ref{Eq6}) for the same time period and four rigidity values used in Figure \ref{fig10} for protons (red triangles) and electrons (blue circles). Reversal period is gray shaded.
Panel (b): Similar to (a) now for the perpendicular MFP in the polar direction (see Equation \ref{Eq7}).
\label{fig11}}
\end{figure*}

In Figure \ref{fig11}(a,b) the variations in the perpendicular MFP in the radial (left panels) and polar direction (right panels) are shown over time, from May 2011 to May 2015, for the same rigidity values used in Figure \ref{fig10}, again for protons (red triangles) and electrons (blue circles). Polarity reversal phase is gray shaded.
The values for protons are much higher initially, before the reversal period, at lower rigidity (0.5 GV and 1.0 GV) compared to values for electrons but the differences gradually and systematically reduce towards the end of the polarity reversal to become somewhat larger again after the reversal period. At these lower rigidity values the MFPs gradually decrease during the reversal period, increasing slightly after the reversal. 
At 3.0 GV the differences become insignificant but the MFP tends to be somewhat higher for electrons than protons in 2012, clearly more so at 10.0 GV. Only in 2015 when the MFPs start to increase as solar activity starts declining, do difference reappear with the MFPs for protons again larger than for electrons.
 In 2011 May, the value of $c_{2\parallel}$ in Equation \ref{Eq5} for protons and electrons is 1.53 and 1.98, respectively, but 1.90 and 1.95 respectively in 2015 May. These trends are all required in order to reproduce observations over time. The value of $c_{2\perp \theta}$ in Equation \ref{Eq7} is kept as 65\% of $c_{2\parallel}$ for both protons and electrons. But the value of $c_{2\perp r}$ in Equation \ref{Eq6} is scaled differently, with its value changing between 60-80 \% of $c_{2\parallel}$. 

\begin{figure*}
\gridline{\fig{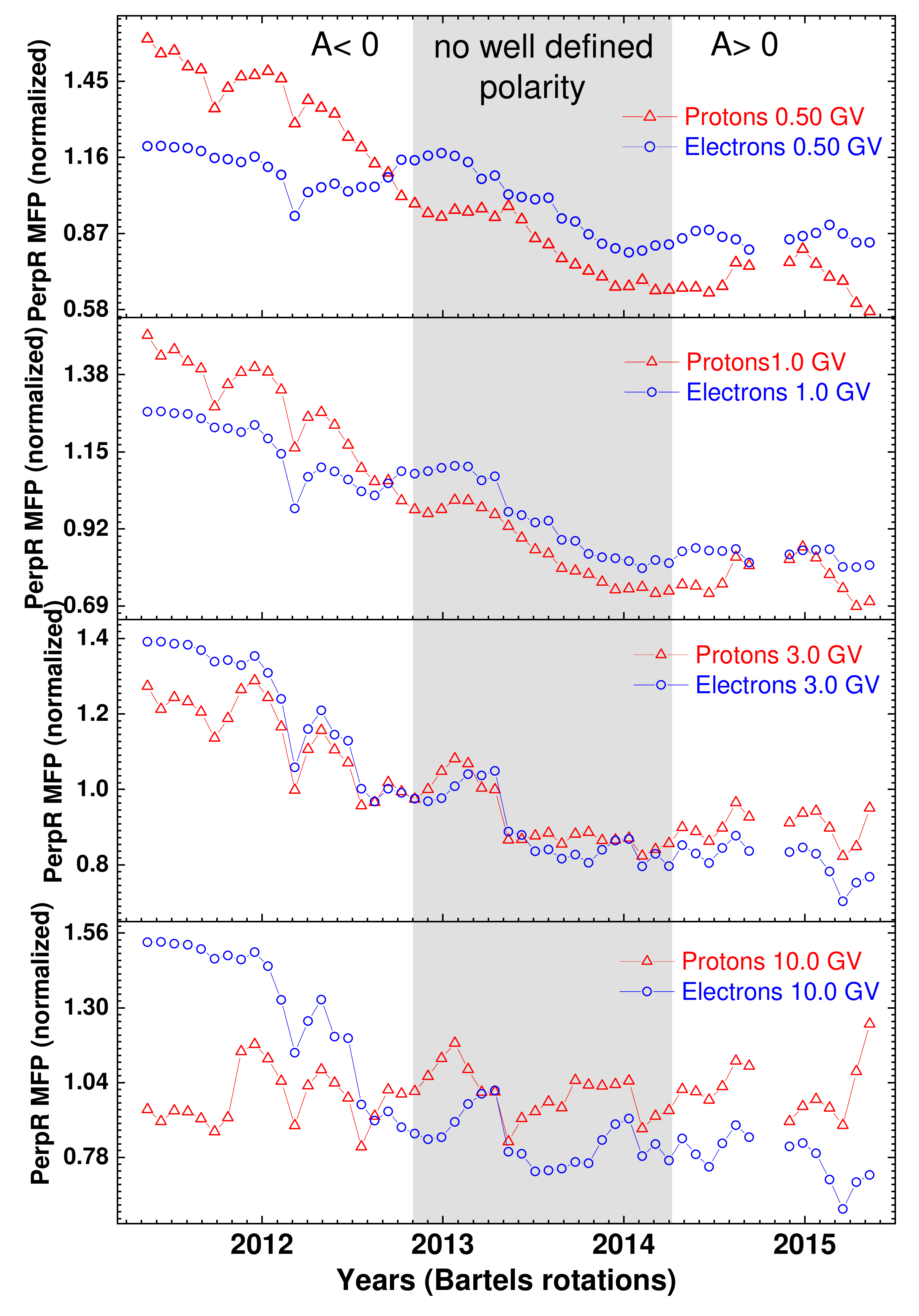}{0.5\textwidth}{(a)}
          \fig{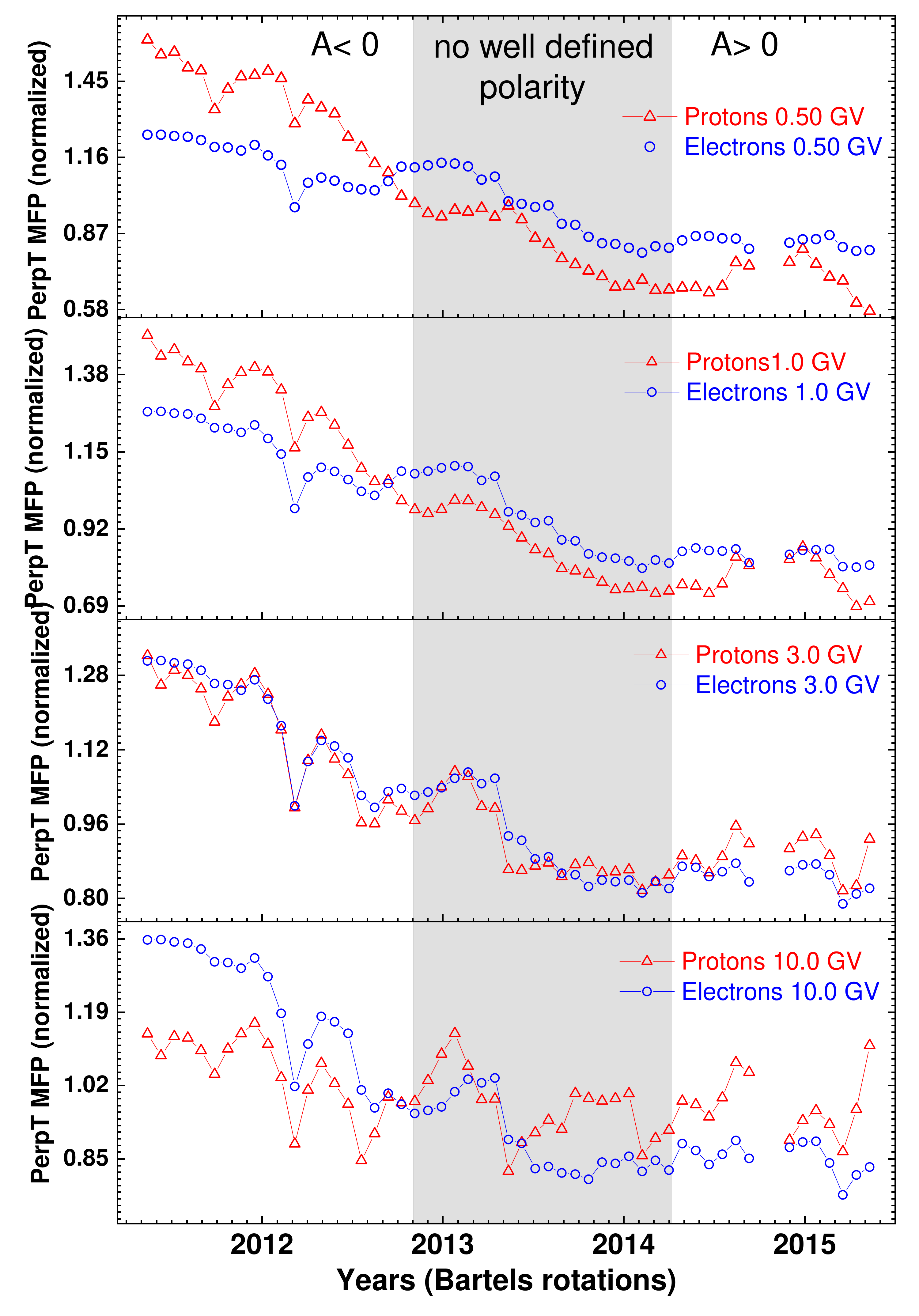}{0.5\textwidth}{(b)}
                }
\caption{Panel (a): Normalized perpendicular MFP in the radial direction for the same period and rigidity values used in Figure \ref{fig11} for protons (red triangles) and electrons (blue circles). 
Panel (b): Similar to (a) now for the normalized perpendicular MFP in the polar direction.
\label{fig12}}
\end{figure*}

For illustrative purposes, Figure \ref{fig12} show the normalized perpendicular MFP variations in the radial direction and in the polar direction, respectively, for the same time period and four rigidity values used in Figure \ref{fig11}, for protons (red triangles) and electrons (blue circles). Their variation shows resemblance with the parallel MFPs at all rigidity values but display a somewhat different rigidity trend. 
\begin{figure*}
\gridline{\fig{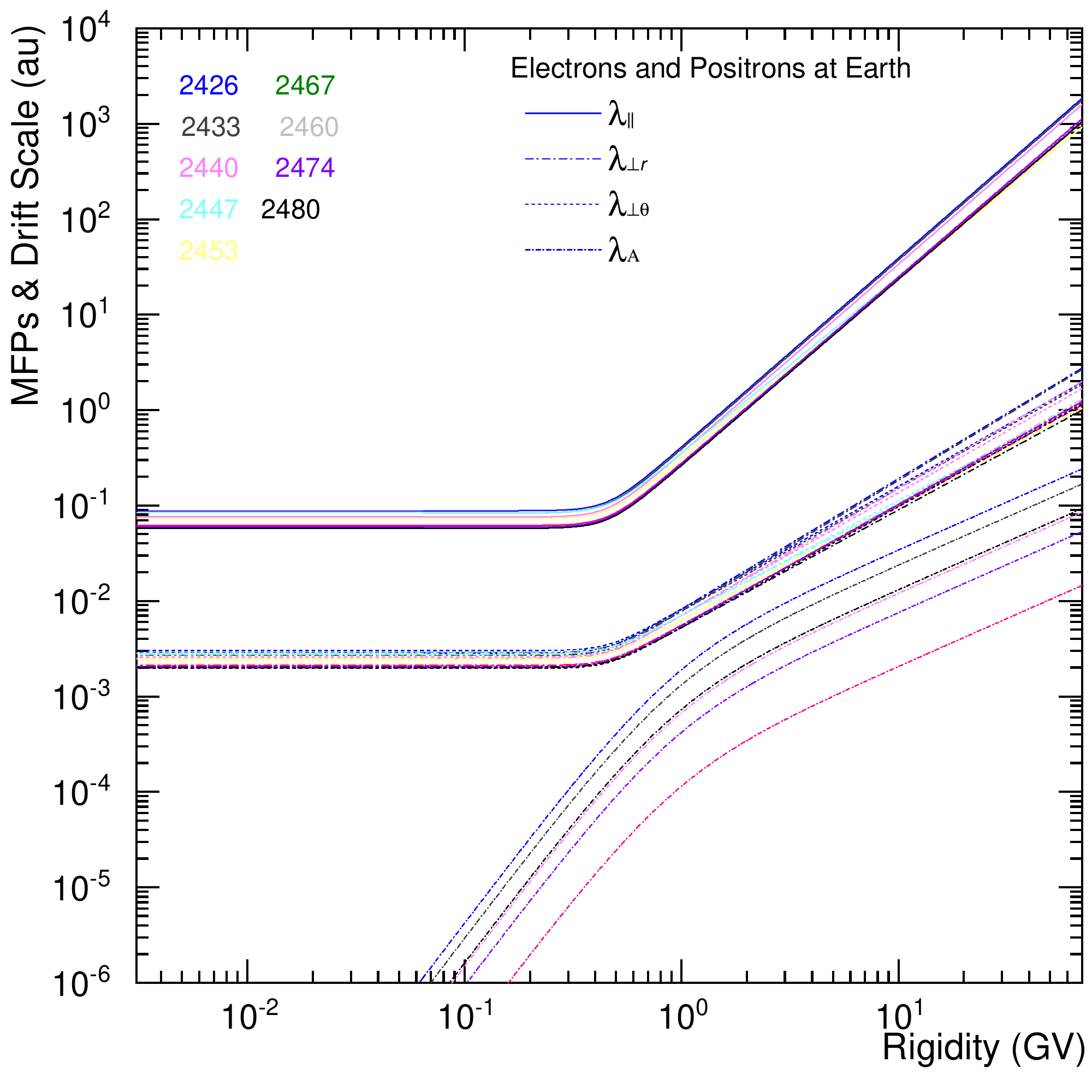}{0.5\textwidth}{(a)}
          \fig{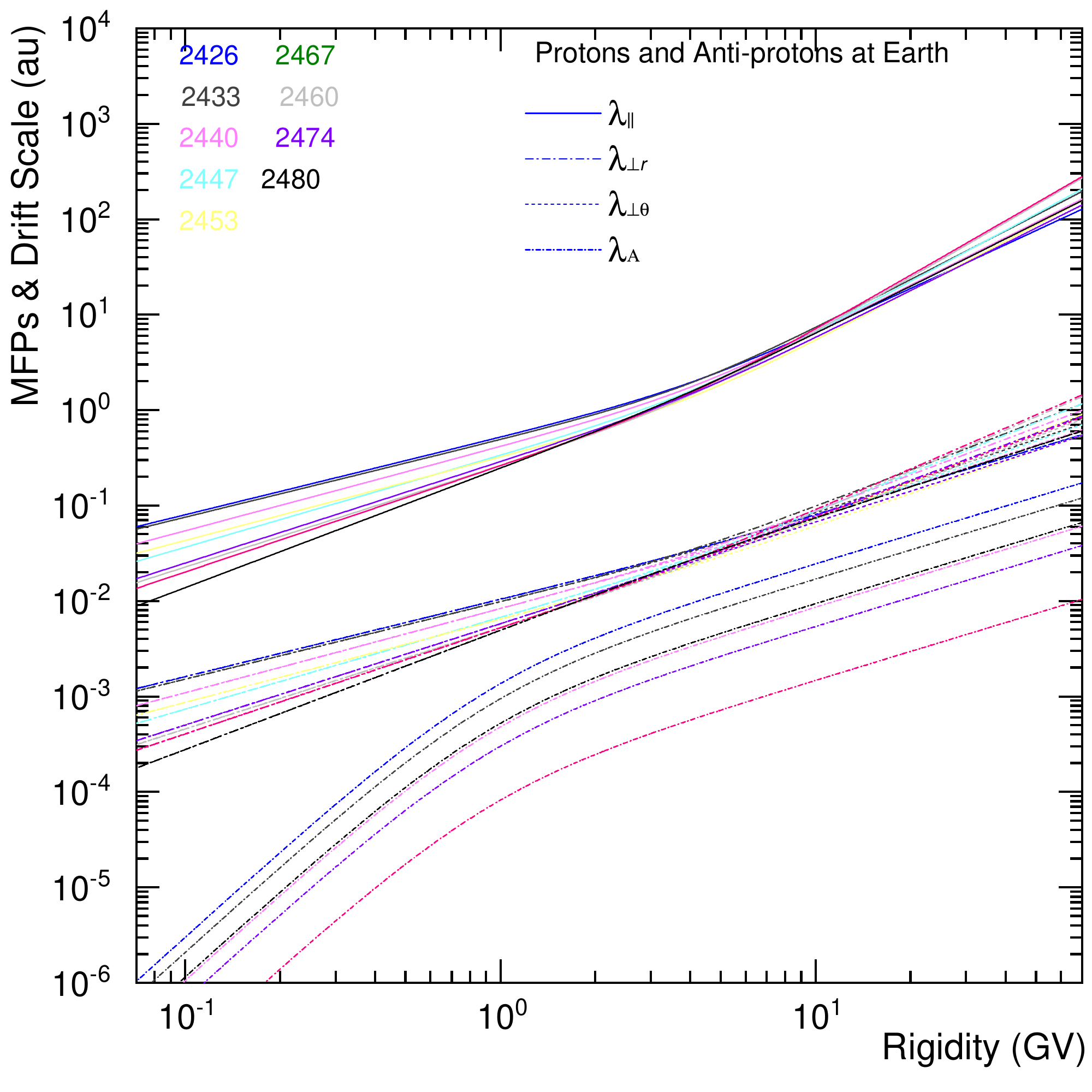}{0.5\textwidth}{(b)}
                 }     
\caption{The rigidity dependence of the three MFPs and drift scale at Earth for electrons and positrons in Panel (a) and for protons and antiprotons in Panel (b), and how they vary from 2011 to 2015 with values for only 9 BRs plotted as indicated (different colours). 
\label{fig13}}
\end{figure*}

Figure \ref{fig13} displays the rigidity dependence of all the MFPs and drift scale at the Earth for (a) electrons and positrons and (b) protons and antiprotons. Values for only 9 BRs are plotted (i.e. it corresponds to $\sim$6-months intervals). Note that for BRs 2446 - 2465 the drift scale was zero over the entire rigidity range as discussed above. The value of $P_{k}$, in Equations \ref{Eq5}, \ref{Eq6} \& \ref{Eq7}, is varied between 0.50-0.60 GV and 2.50-4.0 GV for electrons and protons, respectively. The rigidity dependence of $\lambda_{\parallel}$, $\lambda_{\perp r}$ \& $\lambda_{\perp \theta}$ are different above $P_{k}$. The value of $c_{1} = 0 $ in Equations \ref{Eq5}, \ref{Eq6} \& \ref{Eq7} for electrons but varies between 0.78 to 1.20 for protons. 
What is shown in this figure, is the essence of what is used in the model for the rigidity and time dependence of the diffusion coefficients and the drift coefficient.  

\section{Discussions} \label{sec:discussions}

First, we want to draw attention to the total modulation in general, that is, the full decrease in GCRs from the HP, where the vLIS's are specified, to the Earth as a function of kinetic energy or rigidity, as predicted by the model and depicted in Figure \ref{fig5}. This total modulation differs significantly for GCRs since it clearly depends on the intensity values and the shape of the vLIS specified at the HP for a given type of GCR particle. Of course, it also depends on the modulation volume determined by where the HP is located. This is kept unchanged at 122 au in the model although the width of the heliosheath changes somewhat with solar activity, the effects of which are not explored here; for such a study, see \cite{2016SoPh..291.2181V}. Note how vastly different the electron vLIS is compared to positrons, and the proton vLIS from that for antiprotons. And, subsequently how vastly different the total modulation is, with the total modulation for antiprotons especially been quite peculiar. These specific features and insight of the solar modulation of GCRs can only be gained by applying numerical models. 
In this context, see also the discussions by \citet{2019ApJ...878...59B,2021PAN....84.1121B}, on the various GCR nuclei by \citet{ 2020ApJ...889..167B}, on protons and helium by \citet{2019ApJ...871..253C,2021ApJS..257...48S}, on helium isotopes by \citet{2022AdSpR..69.2330N} and for an overview \citet{2021Potgieter}. 

The next general point of discussion is what exactly is responsible for this total modulation? Of course, the answer is contained in the processes described by the TPE which is solved numerically based on various motivated assumptions for the physical mechanisms as explained above. The difference in the role of the various physical processes follows from Figure \ref{fig5}, for example, note how differently electrons and positrons are modulated with respect to each other and how different this is at low rigidity compared to protons (and antiprotons). 
Note also the differences between these modulated spectra for BR 2426 (May 15-June 11, 2011) compared to BR 2480 (May 12-Jun 08, 2015), before and after the reversal period, with different levels of solar activity, and for which the GCR drift directions are changed. All these aspects contribute to why antiproton modulation show only a 25\% change compared to the other types of GCRs. The peculiar features of antiproton modulation indicates that not even low solar activity can bring significantly more antiprotons to the inner heliosphere.

Of interest is the simulated antiproton to proton flux ratio ($\frac{\bar p}{p}$) at the Earth which seems to be in the order of 10$^{-5}$, more specifically, in May 2011 at 1.0 - 1.16 GV, $\frac{\bar p}{p}$ = 1.062$\times$10$^{-5}$, increasing up to 1.786$\times$10$^{-5}$ by March 2014, then decreasing to 1.29$\times$10$^{-5}$ by May 2015; at 2.67 - 2.97 GV in May 2011, $\frac{\bar p}{p}$ = 5.80$\times$10$^{-5}$, to become 6.83$\times$10$^{-5}$ by March 2014 and 6.21$\times$10$^{-5}$ by Jan 2015, 6.51$\times$10$^{-5}$ by May 2015. At 1 GV this flux ratio shows $\sim$60\% increase until March 2014, but at 3 GV this increase was only $\sim$17\%. At 10.0 GV, in May 2011, $\frac{\bar p}{p}$ = 17.4$\times$10$^{-5}$ changing to  18.5$\times$10$^{-5}$ during the reversal phase, then remaining almost the same, but after the reversal it shows an increase up to 20.0$\times$10$^{-5}$ by May 2015; the flux ratio thus shows $\sim$7\% increase until March 2014. 

In order to follow the trends in GCR observations, our modeling indicates that protons and positrons are modulating stronger during the A$<$0 polarity cycles, in contrast to electrons which are modulated stronger during the A$>$0 cycle. This directly relates to the change in drift direction in the model and is as such convincing evidence of GCR drift effects; see Figure \ref{fig7}. It also follows that since protons and positrons are both positively charged, no drift related effects are evident but it should be noted that at lower rigidity (not shown) large differences will come forward because of the manner adiabatic energy losses occur for these type of GCRs. 

Next, we come to the physics contained in our modeling approach as depicted in Figures \ref{fig10} to \ref{fig13} and how all of this adjusts in detail with every BR as solar activity changes over the particular period studied here. 
The most important aspect in this context, is what is shown in Figure \ref{fig13} for the rigidity dependence of the various MFPs and how this changes for the BRs as indicated; the period is 2011 to 2015. 
The changes in the rigidity slopes in this figure, is obtained through changing the values of the power indices in Equations (\ref {Eq5}) and (\ref {Eq6}).  
For example, in the case of proton and antiproton modulation, the value of c$_{1}$ is $\sim$ 0.80 before the polarity reversal, changing to $\sim$ 1.10 by the end of polarity reversal, and then after the reversal to $\sim$ 1.0, until May 2015, except for the BR 2480; when it is 1.19.  
We find that changing $c_{\perp r}$ over time assists to obtain better reproduction of spectra at higher rigidity. 

Importantly, we confirm, as we reported in our previous studies, that during the polarity reversal phase, the drift scale is effectively zero. 
The rate of decrease of the drift scale before the reversal period is steeper than the rate of increase after the reversal period. All the MFPs keep declining during the reversal period to reach the lowest values at the end of the reversal period. Their recovery are delayed after the reversal period because of the HMF magnitude reached the largest values after the reversal period. Consequently, the recovery of GCRs are also delayed as depicted in Figures \ref{fig1} and \ref{fig2} but differently for electrons because of the particle drift coming back into play. It follows from Figure \ref{fig2} that the relative modulation until the beginning of the reversal period is larger for positrons than for electrons at $\sim$ 1.0 GV and from Figure \ref{fig7}(c) that this is also the case for protons compared to antiprotons. This again relates to the drift patterns of these oppositely charged particles.

In all our published numerical modeling, we assume these heliospheric MFPs to be identical for electrons and positrons, and also identical for protons and antiprotons (in fact also for helium and all other GCR nuclei), because there are no reason, or any arguments, why they should differ as a function of rigidity. But, there are meaningful differences between the proton and electron MFPs, with noteworthy results on  how the MFPs change with time (with solar activity), how their rigidity dependence changes with time and how the turn-over rigidity, $P_{k}$, changes with time. 
When the value of the MFPs get lower (stronger diffusive, increased modulation), the effects of changing c$_{1}$ becomes easily noticeable. During solar minimum when the MFPs are comparatively larger (less modulation), the effect of changing c$_{1}$ is masked. 
As such an interesting point: despite its complexity, the polarity reversal period is a good phase to study modulation effects; not only is the drift scale effectively zero, also the diffusion coefficients are at their lowest levels which further expose the influence and details of the physical processes. 

Our simulations indicate that $\lambda_{\parallel}$ is much higher ($\sim$ 3 times) for protons compared to electrons at lower (0.5 GV) rigidity in 2011, but at 10.0 GV this is the opposite i.e., $\lambda_{\parallel}$ is much higher higher ($\sim$ 3 times) for electrons compared to protons. This difference decreases to a minimum by the end of polarity reversal with similar trends in $\lambda_{\perp r}$ \& $\lambda_{\perp \theta}$. The difference is that for $\lambda_{\parallel}$ the shifting happens between 1-3 GV, but for $\lambda_{\perp r}$ \& $\lambda_{\perp \theta}$ the shifting happens around 3 GV. The relates to the assumption that the rigidity dependence of $\lambda_{\parallel}$ is not identical to that for $\lambda_{\perp r}$ \& $\lambda_{\perp \theta}$. 
Also note that the rigidity dependence of the drift scale is quite different from that for the various MFPs.

In our earlier numerical studies, the value of $P_{k}$ for protons was fixed at exactly 4.0 GV, but here we find that it should change with time; varying between 2.50 to 4.0 GV in order to reproduce observed spectra better at higher rigidity (5-10 GV). Introducing this type of time variation has a slightly different impact during A$<$0 cycles compared to A$>$0. 
In this context, see also the elaborate simulations done by \cite{2021ApJS..257...48S} using a different type of numerical model and approach than what we report here. For electrons, where $P_{k} = $0.4 GV, changing this value seems required to a lesser degree.

The ratio of the antiproton flux with the three other GCR species shows interesting variations; the highest ratio with protons and positrons occurs around the end of the polarity reversal at lower rigidity, decreasing after the reversal. The ratio with electrons continues to increase up to May 2015.
Our simulations show that at lower rigidity the antiproton to proton and antiproton to positron ratio signify the reversal of drift direction after the reversal. Antiproton to electron ratio indicates that even though the majority of electrons are entering through the polar region until the reversal, the electron flux decreases as solar activity increases since 2011, but the antiproton flux is already low so that a change in solar activity is relatively less effective, causing only a slight increase in the flux ratio. This trend continues even after the polarity reversal and we expect that it may stay a little longer, then start to reverse when drift get stronger (in May 2015 only 38\%).

It is worth mentioning that during the polarity reversal period $\sim$ 40 ICMEs were reported, most of them produced geomagnetic disturbances, but interestingly not any meaningful fluctuations in AMS observations, not even at 1.0 GV.

\section{Summary and Conclusions} \label{sec:summary}

The primary objective of this study is to understand at a deeper level how the major modulation processes, especially particle drift, vary from before to after the HMF polarity reversal period, that is, from a negative to positive polarity phase and through a period of almost 1.5 years of an uncertain polarity.

A well-established, sophisticated 3D numerical modulation model is applied to simulate and in the process also reproduce in detail AMS observations of galactic proton, electron and positron fluxes, and simulating the corresponding antiproton flux for the period May 2011 to May 2015. This covers the reversal period of Solar Cycle 24, and is done for BR averaged observations and simulations for the whole period under investigation. As before, the tilt angle of the HCS and magnitude of the HMF measured at the Earth are used as proxies for solar activity and utilized as time-varying input for the numerical model. Along with running averages for these values for the studied period, the values of the diffusion and drift coefficients are changed in order to simulate and effectively reproduce the published AMS02 observations.  

All the vLISs utilized here are initially computed by applying the GALPROP code, then adjusted based on the Voyager 1 \& Voyager 2 observations at lower energy, below $\sim$ 10 MeV, and the PAMELA and AMS observations at higher energy. See the overview by \citet{2021Potgieter}.
Concerning the modulated spectra at the Earth, our model gives the peak intensity for electrons and positrons just below 1 GV, for protons just above 1 GV, but for antiprotons the peak is around 3 -4 GV.
According to \cite{2022AdSpR..69.2330N}, the peak intensity for GCR helium is between 1 and 2 GV. Note that the rigidity values of these peaks in spectral intensity do change with solar activity, always to a higher rigidity for increased modulation.

The modeling results together with observations show for 1-3 GV that the lowest fluxes for the GCRs investigated, occur at the end of the reversal period but while the proton and positron fluxes then started to recovery, the electrons, and simulated antiprotons, stayed at these low values for about a year longer. According to the model, this indicates the effect of the changing drift directions.

We have illustrated that our 3D model, with the physics it contains, is able to reproduce AMS observations for the rigidity range of 1-3 GV to a large degree, followed by simulating galactic antiprotons for the mentioned period. The difference in how these fluxes change after the polarity reversal caused by the change of the particle drift direction is noticeable even at 3 GV. The flux variation over time of electrons and antiprotons are similar, as is the flux variation of protons and positrons. 
Applying of the changing drift patterns in the model, the relative electron flux at 1-3 GV is higher during the A$<$0 phase compared to the other considered GCRs, but after the reversal, the electron flux is lower. The variation in the simulated antiproton flux, from the normal, is less compared to the other considered GCRs for all three modulation phases (A$<$0 cycle, polarity reversal period, and A$>$0 cycle). As expected, the largest flux variations occur at the lowest rigidity, getting significantly reduced as rigidity is increased.

The time variations simulated for the flux ratio for electrons to protons and for electrons to positrons confirm the effect of the reversal of the drift directions after the HMF reversal. The time variations for the antiproton to proton and antiproton to positron ratio draw the same conclusion.  
The way in which the ratio of antiprotons to electrons varies, confirms that both these GCRs experience  the same drift directions during all polarity phases.

Our modeling indicates that the global drift of charged particles in the heliosphere becomes negligible during the HMF polarity reversal period which is consistent to what \cite{2005AnGeo..23.1061N} reported when modeling Ulysses observations of electrons and protons at 2.5 GV during the short reversal period in 2000. The drift scale needs to be adjusted with time in order to reach and keep at its lowest level (essentially zero) during the HMF reversal phase; the rate of change in the drift scale is different for the period before the HMF reversal than afterwards. Consequently, the diffusion process is dominant in modulating the GCR particles during the polarity reversal but with adiabatic energy changes always overwhelmingly present in the case of protons and antiprotons at lower rigidity; see Figure \ref{fig3}.

In contrast to their large values during solar minimum activity, all MFPs during the polarity reversal give relatively low values which as such expose the relative contributions, effects and level of the different modulation processes from minimum to maximum solar activity. The modeling indicates that the drift scale starts recovering after the polarity reversal, but the MFPs keep decreasing or remain unchanged for some period after the polarity reversal. It is important to keep in mind that the modulation of GCRs at the Earth depends on what happens between 1 au and the HP at 122 au and that there is a time-lag between cosmic ray modulation and solar variability \citep[see, e.g.][and references therein]{2022SoPh..297...38K}.

A next phase of our modeling on charge-sign dependence will be focused on the period after May 2015 to include the solar minimum modulation period around 2020 and what is predicted for this time; see e.g. \citet{2021AdSpR..68.2953K} and references therein. Comprehensive modeling of   helium in comparison with protons was reported by \cite{2019ApJ...871..253C,2021ApJS..257...48S} and for the two Helium isotopes by \citet{2022AdSpR..69.2330N}.
Incorporating also observations at lower rigidity \citep[see, e.g.][and references therein]{2019A&A...625A.153M,2020ApJ...901....8B} than what AMS can provide, is considered essential for further modeling studies. 



\begin{acknowledgments}

The present work is supported by the Shandong Institute of Advanced Technology (SDIAT) and NSFC Project U2106201. MDN acknowledges the SA National Research Foundation (NRF) for partial financial support under the Joint Science and Technology Research Collaboration between SA and Russia (Grant no:118915).
The authors wish to thank the GALPROP developers and their funding bodies for access to and use of the GALPROPWebRun service. We acknowledge the use of HCS tilt angles and Solar Polar Field data from the Wilcox Solar Observatory's website \url{http://wso.stanford.edu}, HMF and Sunspot Numbers data from NASA's OMNIWEB data interfaces \url{http://omniweb.gsfc.nasa.gov} and \url{https://www.sidc.be/silso/home}, and Oulu NM records from \url{https://cosmicrays.oulu.fi/}. 
\end{acknowledgments}




\end{document}